\documentclass[preprint,preprintnumbers,nofootinbib,aps,10pt]{revtex4-1}
 \usepackage{amsmath,amssymb,bm,epsfig}
 \usepackage{color}
 \usepackage{natbib}
 \usepackage{hyperref}
 \usepackage{ulem}
 \usepackage{graphicx}
 \usepackage{xcolor,colortbl}
 \usepackage{pifont}

\def \beq{\begin{equation}}
\def \eeq{\end{equation}}
\def \beqa{\begin{eqnarray}}
\def \eeqa{\end{eqnarray}}
\def \l{\left(}
\def \r{\right)}

\newcommand{\nn}{\nonumber}

\usepackage{xspace}

\newcommand{\sNN}{\sqrt{s_{\rm NN}}}
\newcommand{\rhom}{\rho^{\text{max}}}
\newcommand{\tauf}{\tau^{\text{f}}}
\begin{document}

\title{The Freezeout Hypersurface at LHC from particle spectra: Flavor and Centrality Dependence 
}

\author{Sandeep Chatterjee}
\email{sandeepc@vecc.gov.in}
\affiliation{Theoretical Physics Division, 
Variable Energy Cyclotron Centre, 1/AF Bidhannagar, 
Kolkata, 700064, India}

\author{Bedangadas Mohanty}
\email{bedanga@niser.ac.in}
\affiliation{School of Physical Sciences, National Institute 
of Science Education and Research, Jatni, 752050, India}

\author{Ranbir Singh}
\email{ranbir.singh@niser.ac.in}
\affiliation{School of Physical Sciences, National Institute 
of Science Education and Research, Jatni, 752050, India}

\begin{abstract}
We extract the freezeout hypersurface in Pb-Pb collisions at $\sqrt{s_{\rm NN}}=$ 2760
GeV at the CERN Large Hadron Collider by analysing the data on transverse momentum spectra 
within a unified model for chemical and kinetic freezeout. The study has been done within 
two different schemes of freezeout, single freezeout where all the hadrons freezeout together 
versus double freezeout where those hadrons with non-zero strangeness content have different 
freezeout parameters compared to the non-strange ones. We demonstrate that the data is better 
described within the latter scenario. We obtain a strange freezeout hypersurface which 
is smaller in volume and hotter compared to the non-strange freezeout hypersurface for all 
centralities with a reduction in $\chi^2/N_{df}$ around $40\%$. We observe from the extracted 
parameters that the ratio of the transverse size to the freezeout proper time is invariant under 
expansion from the strange to the non-strange freezeout surfaces across all centralities. Moreover, 
except for the most peripheral bins, the ratio of the non-strange and strange freezeout proper 
times is close to $1.3$.

PACS numbers:12.38.Mh, 25.75.-q, 25.75.Dw, 25.75.Ld

\end{abstract}
\maketitle

\section{Introduction}\label{sec.intro}

The characterisation of the freezeout hypersurface (FH) is a major goal in heavy ion 
phenomenology. A well determined FH allows one to compute the hadronic contribution from the 
surface of last scattering to most soft physics observables. This then forms the standard 
baseline expectation to be compared with data to open the window to physics from times earlier 
than freezeout. Hence, a reliable baseline is essential to any efforts to detect signatures of
the quark gluon plasma (QGP) phase as well as the physics of the Quantum Chromodynamics (QCD) 
critical point~\cite{Borsanyi:2014ewa,Alba:2014eba,Bazavov:2012vg}. The thermal parameters at 
the time of chemical freezeout (CFO) where inelastic collisions cease are determined from the 
data on hadron multiplicities~\cite{BraunMunzinger:1995bp,Yen:1998pa,Becattini:2000jw,Andronic:2005yp}. 
The particle transverse momentum spectra gets fixed only when the elastic collisions cease and their 
study leads to understanding of the fireball conditions at the time of kinetic freezeout (KFO). 
There have been suggestions of sudden freezeout~\cite{Letessier:1994cn,Csorgo:1994dd,Csernai:1995zn}. 
However the earlier analysis of spectra data based on the blast wave model~\cite{Schnedermann:1993ws} at 
the SPS and RHIC energies found the KFO temperature to be around 100-130 
MeV~\cite{Afanasiev:2002fk,BurwardHoy:2002xu}. This is significantly lower than the CFO temperature 
which is around 150-160 MeV at the same energies~\cite{Andronic:2005yp} hence suggesting a delayed 
KFO compared to CFO. Later efforts where the complete two and three body resonance decays were 
taken into account showed that the RHIC data could be explained 
by simultaneous CFO and KFO through a common set of thermal and geometric parameters that describes 
the FH~\cite{Broniowski:2001we,Broniowski:2001uk}. This standard single freezeout scheme (1FO) invovles 
four free parameters which are determined by fits to data of particle ratios and $p_T$ spectra: two 
thermal parameters namely temperature $T$ and the baryon chemical potential $\mu_B$ and two geometric 
parameters to parametrise the FH.

At the LHC energy of $\sNN=2760$ GeV, 1FO scheme with three parameters (at LHC energies $\mu_B$ 
can be set to zero) failed to describe the transverse 
momentum spectra and this was reported as the proton anomaly~\cite{Rybczynski:2012ed}. Even $p_T$ integrated 
yields can not be satisfactorily described within this scheme~\cite{Stachel:2013zma}. 
This has prompted a lot of interest in the physics of freezeout and various alternatives to describe 
the yields have been suggested. All these alternative schemes introduce additional free parameters that are 
rquired to be tuned to fit the data. In Refs.~\cite{Steinheimer:2012rd,Becattini:2012xb,Becattini:2012sq}, 
hadronization followed by late stage baryon-antibaryon annihilations were taken into account through a 
microscopic UrQMD approach that resulted in better agreement with data. A study on the centrality dependence of the 
freezeout parameters by investigating the particle yields at $\sNN=$ 2760 GeV within the above 
scheme was done in Ref.~\cite{Becattini:2014hla}. In another approach, effects of nonequilibrium physics 
were incorporated by introducing additional light and strange quark phase space factors, the values of 
which were tuned to describe the particle yields~\cite{Petran:2013lja}. Thus, this scheme required five 
free parameters. 
Subsequently, the centrality dependence of the freezeout parameters at LHC within the nonequlibrium 
chemical freezeout picture (NFO)  by analysing the $p_T$ spectra was done~\cite{Begun:2013nga,Begun:2014rsa}. 
However, it was also realised in Ref.~\cite{Begun:2014rsa} that for a consistent description of the spectra 
of the strange hadrons 
like $\Lambda$, $\Xi$ and $\Omega$ within the NFO picture one required an additional 
geometric parameter and this was interpreted as early freezeout for the strange hadrons as have been argued in 
Ref.~\cite{Chatterjee:2013yga,Bugaev:2013sfa} in a equilibrium scenario. Thus, a consistent description 
of the $p_T$ spectra for all the observed hadrons could be done in the NFO scenario with 
six free parameters~\cite{Begun:2014rsa}.
While both these schemes involve aspects of nonequilibrium physics, there has been yet another proposal 
completely based on equilibrium thermodynamics where hadrons with non-zero strangeness content freezeout 
earlier compared to those with vanishing strangeness (2FO)~\cite{Chatterjee:2013yga,Bugaev:2013sfa}. The 
production of light nuclei and hypernuclei in 2FO has been recently studied~\cite{Chatterjee:2014ysa}. 
At $\sNN=200$ GeV, 2FO successfully describes ratios of strange to non-strange nuclei~\cite{Chatterjee:2014ysa} 
which has so far been missing in studies based on 1FO~\cite{Andronic:2010qu,Cleymans:2011pe}. In this 
paper we will investigate whether 2FO with six free parameters (three parameters each for the non-strange 
and strange FH) provides a better description of the $p_T$ spectra data  compared to 1FO. This will also 
clarify whether the NFO scenario which also requires six free parameters is the only picture that describes 
the LHC data or 2FO is a viable alternative scenario. We also further investigate the systematics of the 
extracted freezeout parameters and find interesting correlations between them. These correlations can be 
used as constraints, reducing the number of free parameters to five and in case of $\l0-40\r\%$ centralities to 
even four.

We compare the particle spectra description as obtained in 1FO and 2FO schemes of freezeout. Here 
is a brief outline of the paper. In Section~\ref{sec.model}, we describe the 1FO scheme that we use which is 
also implemented in the Monte Carlo event generator 
THERMINATOR~\cite{Kisiel:2005hn,Chojnacki:2011hb}. We then discuss its extension in the 2FO case. 
In Section~\ref{sec.results}, we analyse the $p_T$ spectra data and make a comparative study of 1FO 
and 2FO schemes. We have investigated the centrality dependence of the FH by analysing the data in 
six centrality bins: $\l00-05\r \%$, $\l05-10\r \%$, $\l10-20\r \%$, $\l20-40\r \%$, $\l40-60\r \%$ 
and $\l60-80\r \%$ which we will refer to as centrality bin number 1, 2, 3, 4, 5 and 6 
respectively. Finally in Section~\ref{sec.summary} we summarise our results and conclude.

\section{Model}\label{sec.model}

The basic equation for the study of the transverse momentum spectra and the FH is the Cooper-Frye 
formula that gives the rapidity and transverse momentum distribution of the produced hadrons at 
the surface of last scattering,
\beq
\frac{d^2N}{dy_pp_Tdp_T}=\int d\Sigma\cdot pf\l p\cdot u,T,\mu_B,\mu_Q,\mu_S\r
\label{eq.cooper}
\eeq
where the integration is over the FH of which $d\Sigma^{\mu}$ is a differential element. $u^{\mu}$ 
is the hydrodynamic 4-velocity of the fireball at the FH. Different models of freezeout differ in 
the choice of the FH and parametrization of $u^{\mu}$. In this paper we follow the boost invariance 
prescription and the freezeout takes place at a fixed invariant time $\tauf$. Thus a natural 
parametrization of the FH is
\beq
\l\tauf\r^2=t^2-x^2-y^2-z^2
\label{eq.1FO}
\eeq
where $\tauf$ is fixed by data. We use the following parametrisation for $u^{\mu}$ at 
$x^{\mu} = (t, x, y, z)$
\beq
u^{\mu}=x^{\mu}/\tauf
\label{eq.velocity}
\eeq

The distribution function $f\l p\cdot u, T, \mu_B, \mu_Q, \mu_S\r$ comprise of primary hadrons as 
well as secondary ones that arise due to decays of the unstable resonances. The decays and the 
momentum distribution thereof are taken care of by THERMINATOR~\cite{Kisiel:2005hn,Chojnacki:2011hb}. 
The main distribution that goes into $f$ is the primordial distribution of the $i^{\text{th}}$ hadron
\beq
f_p^i\l p\cdot u, T, \mu_B, \mu_Q, \mu_S\r=\frac{g_i}{\exp{\left[\l p\cdot u-B_i\mu_B-S_i\mu_S-Q_i\mu_Q\r/T\right]}+a_i}
\label{eq.distribution}
\eeq
where $g_i$ is the degeneracy of the $i^{\text{th}}$ hadron and $a_i = 1\l-1\r$ for baryons (mesons). 
A natural choice of coordinates to describe the FH are space-time rapidity $y_s$, azimuthal angle $\phi$ 
and the distance $\rho$ of the freezeout surface from the beam line given by $\rho=\sqrt{x^2+y^2}$. While 
$y_s$ is integrated from minus infinity to plus infinity and $\phi$ from $0$ to $2\pi$, the integration 
range of $\rho$ is from 0 to $\rhom$, where $\rhom$ is interpreted as the edge of the fireball cylinder.
We note that the $y_s$ dependence of the integrand in Eq.~\ref{eq.cooper} that enters through $p\cdot u$
\beq
p\cdot u=\frac{p_0}{2\tau_f}\sqrt{\tau_f^2+\rho^2}\l e^{y_s}+e^{-y_s}\r-p_x\frac{\rho\cos\phi}{\tau_f}
-p_y\frac{\rho\sin\phi}{\tau_f}-\frac{p_z}{2\tau_f}\sqrt{\tau_f^2+\rho^2}\l e^{y_s}-e^{-y_s}\r
\label{eq.pdotu}
\eeq
for the choice of $u^\mu$ in Eq.~\ref{eq.velocity} ensures convergence of the integral in 
Eq.~\ref{eq.cooper}.

The normalization of the distribution in Eq.~\ref{eq.distribution} is fixed from Eq.~\ref{eq.cooper} by 
comparing with the experimentally measured yields. In Ref.~\cite{Begun:2014rsa} it has been shown that from 
Eq.~\ref{eq.cooper}, the rapidity density of hadron $i$ can be written as
\beq
\frac{dN_i}{dy}=\pi\l\rhom\r^2\tauf n_i\l T,\mu_B,\mu_Q,\mu_S\r
\label{eq.yield}
\eeq
where $n_i\l T,\mu_B,\mu_Q,\mu_S\r=\int \frac{d^3p}{\l2\pi\r^3}{f_p^i}_{LRF}\l p, T, \mu_B, \mu_Q, \mu_S\r$.
Here ${f_p^i}_{LRF}$ is the distribution function $f_p^i$ from Eq.~\ref{eq.distribution} In the local rest 
frame (LRF) ($u_{\mu}=\{1,0,0,0\}$),
\beq
{f^i_p}_{LRF}\l p, T, \mu_B, \mu_Q, \mu_S\r=\frac{g_i}{\exp{\left[\l\sqrt{m_i^2+p^2}-B_i\mu_B-S_i\mu_S-Q_i\mu_Q\r/T\right]}+a_i}.
\label{eq.localdistribution}
\eeq
Thus, as evident from Eq.~\ref{eq.yield}, the model parameters $\tauf$ and $\rhom$ control the normalization 
which have been fixed here by data on hadron spectra.

The above parametrizations of the FH and flow velocity in Eqs.~\ref{eq.1FO} and \ref{eq.velocity} 
and setting the FH edge at a constant $\rho=\rhom$ are strictly true only for central collisions. 
In case of collisions with non-zero impact parameter, we expect the initial state spatial anisotropy 
of the overlap region to leave imprint on the final state FH that will modify $u^{\mu}$ and the FH to 
account for flow effects. However, in this paper we will not compare with data on flow as our main aim 
is to look for evidence of flavor dependence in freezeout from data on $p_T$ spectra. Hence, even for 
non-central collisions we use the above procedure~\cite{Rybczynski:2012ed,Begun:2013nga,Begun:2014rsa}. Thus 
while for central collisions our approach is perfect, in case of peripheral collisions we will be able 
to have only an approximate estimate of the FH. In future we aim to perform an estimation of the FH in 
case of non-central collisions by including the effects of anisotropy and compare with data on flow as
in Ref.~\cite{Broniowski:2002wp}. However, Ref.~\cite{Broniowski:2002wp} keeps the uniform Bjorken flow 
assumption, Eq.~\ref{eq.velocity}, thus the initial shear and vorticity are neglected and these are 
significant in peripheral collisions. So a more realistic approach would be needed for precise modeling.

Thus, in this scheme there are six parameters: $T$, $\mu_B$, $\mu_Q$, $\mu_S$, $\rho_{max}$ and $\tauf$. While $T$, 
$\mu_B$, $\mu_S$ and $\mu_Q$ are extracted from fits to $p_T$ integrated data on hadron multiplicities, the 
remaining two parameters, $\rhom$ and $\tauf$ are extracted from fits to $p_T$ spectra. For the LHC conditions, 
all the chemical potentials can be taken to be zero. Thus in the 1FO scheme we are left with only three parameters 
to be obtained from fit to data. The extension of the 1FO scheme to the 2FO scheme is very simple. In 
the 2FO scheme, hadrons with non-zero strangeness content freezeout earlier compared to the rest. Thus we have 
now two FHs,
\beqa
\l\tauf_s\r^2 &=& t_s^2-x_s^2-y_s^2-z_s^2\\
\l\tauf_{ns}\r^2 &=& t_{ns}^2-x_{ns}^2-y_{ns}^2-z_{ns}^2
\label{eq.2FO}
\eeqa
where the subscripts ’s’ and ’ns’ stand for the strange and non-strange FHs respectively. Eqs.~\ref{eq.distribution} 
and \ref{eq.velocity} are also modified accordingly. Finally, we have six parameters in 2FO: $T_s$ and $T_{ns}$ 
are extracted from fits to multiplicity while $\rhom_s$, $\rhom_{ns}$, $\tauf_s$ and $\tauf_{ns}$ are extracted 
from fits to spectra.

\section{Results}\label{sec.results}

 \begin{table*}[bt]
 \begin{center}
 \begin{tabular}{|c|c|c|c|c|c|c|c|c|c|}
 \hline
  centrality & $T$ & $T_s$ & $T_{ns}$ & $\rhom$ & $\rhom_s$ & $\rhom_{ns}$ &
  $\tauf$ & $\tauf_{s}$ & $\tauf_{ns}$\\
  $\%$ & (MeV) & (MeV) & (MeV) & (fm) & (fm) & (fm) & (fm) & (fm) & (fm) \\
 \hline
  00-05& 155(2) & 164(7) & 149(4) & 13.4 & 11.4 & 14.5 & 9.5 & 7.7 & 10.2 \\
 \hline
  05-10& 155(2) & 166(6) & 149(4) & 12.2 & 10.5 & 13.8 & 8.6 & 7.4 & 9.6 \\ 
 \hline 
  10-20& 156(2) & 162(2) & 146(4) & 11.0 & 10.1 & 13.3 & 7.9 & 7.0 & 9.2 \\
 \hline
  20-40& 162(3) & 163(4) & 147(5) & 8.3 & 8.1 & 10.9 & 6.1 & 6.0 & 7.8 \\
 \hline 
  40-60& 160(3) & 162(4) & 150(5) & 6.1 & 6.0 & 7.3 & 4.8 & 4.7 & 5.6 \\
 \hline
  60-80& 154(3) & 158(3) & 151(6) & 4.2 & 4.0 & 4.5 & 3.9 & 3.4 & 3.9 \\
 \hline
 \end{tabular}
 \end{center}
 \caption{The thermal and geometric freezeout parameters in 1FO vs 2FO at $\sNN=2760$ GeV. The 
 errors in temperatures are estimated from fits to yields while the corresponding errors on $\rhom$ 
 and $\tauf$ are estimated to be around $20\%$ for all centralities.}
 \label{tab.2FOparams}\end{table*}
 
  \begin{figure}[htb]
  \begin{center}
    \scalebox{0.29}{\includegraphics{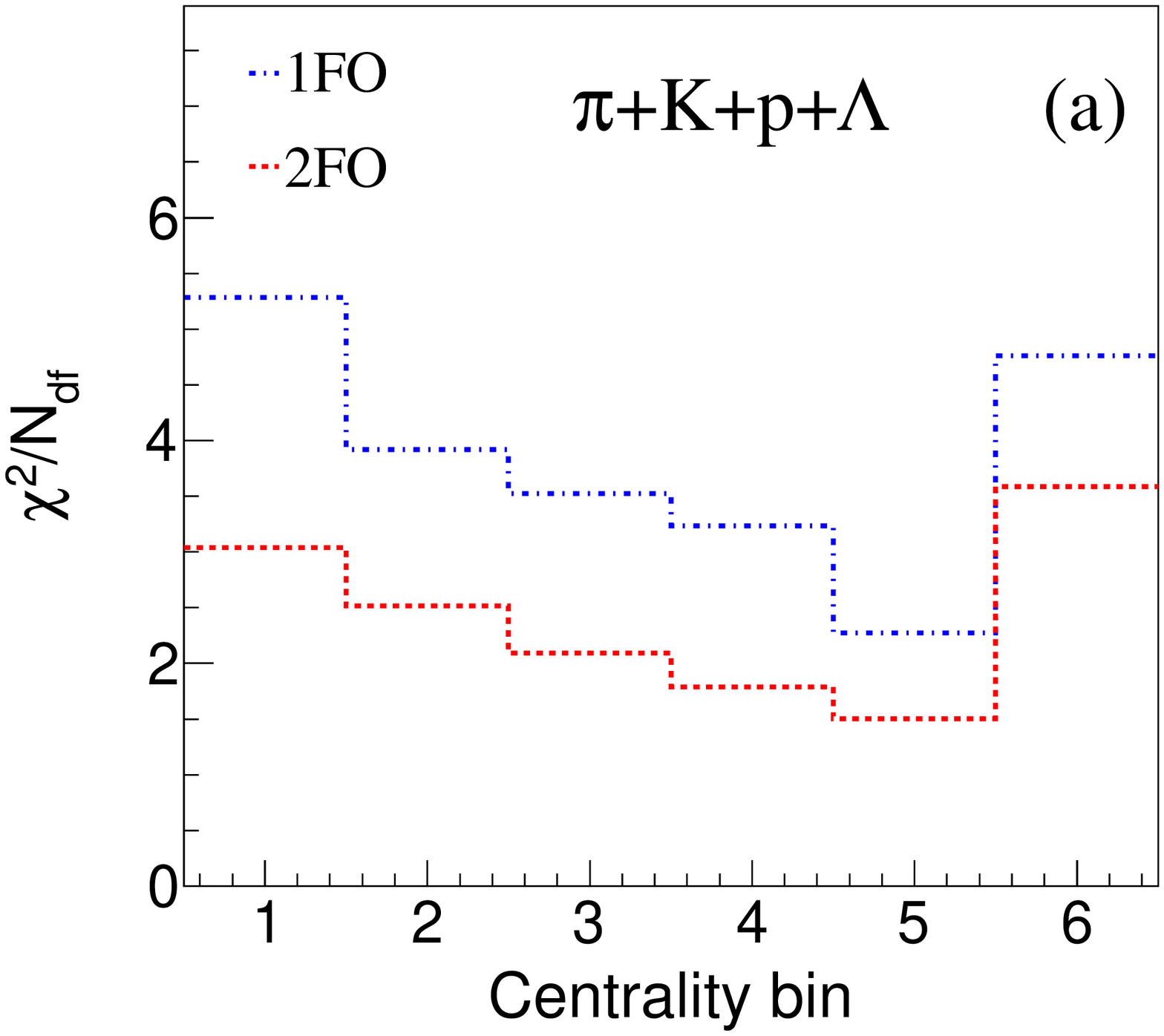}}
    \scalebox{0.29}{\includegraphics{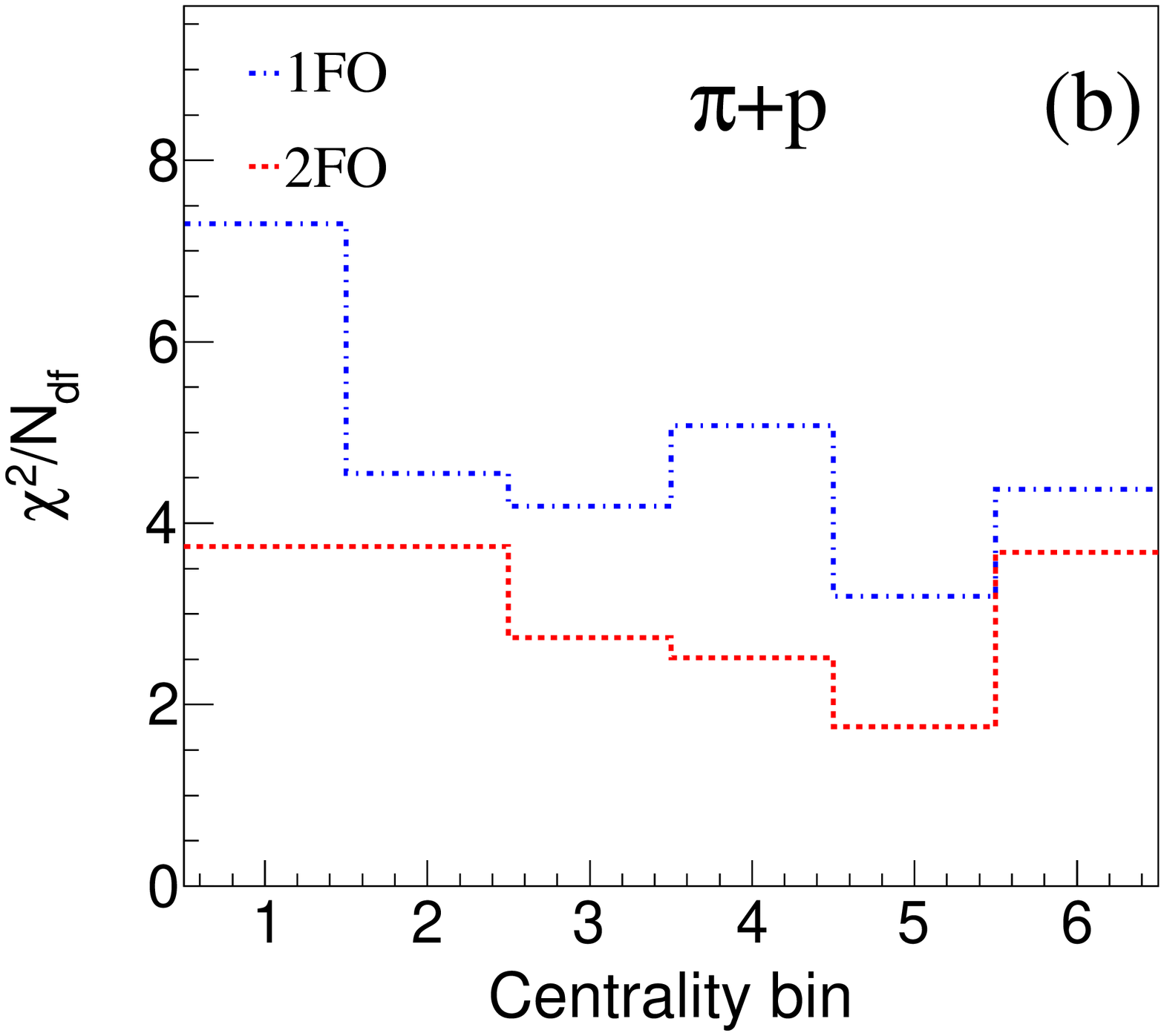}}
    \scalebox{0.29}{\includegraphics{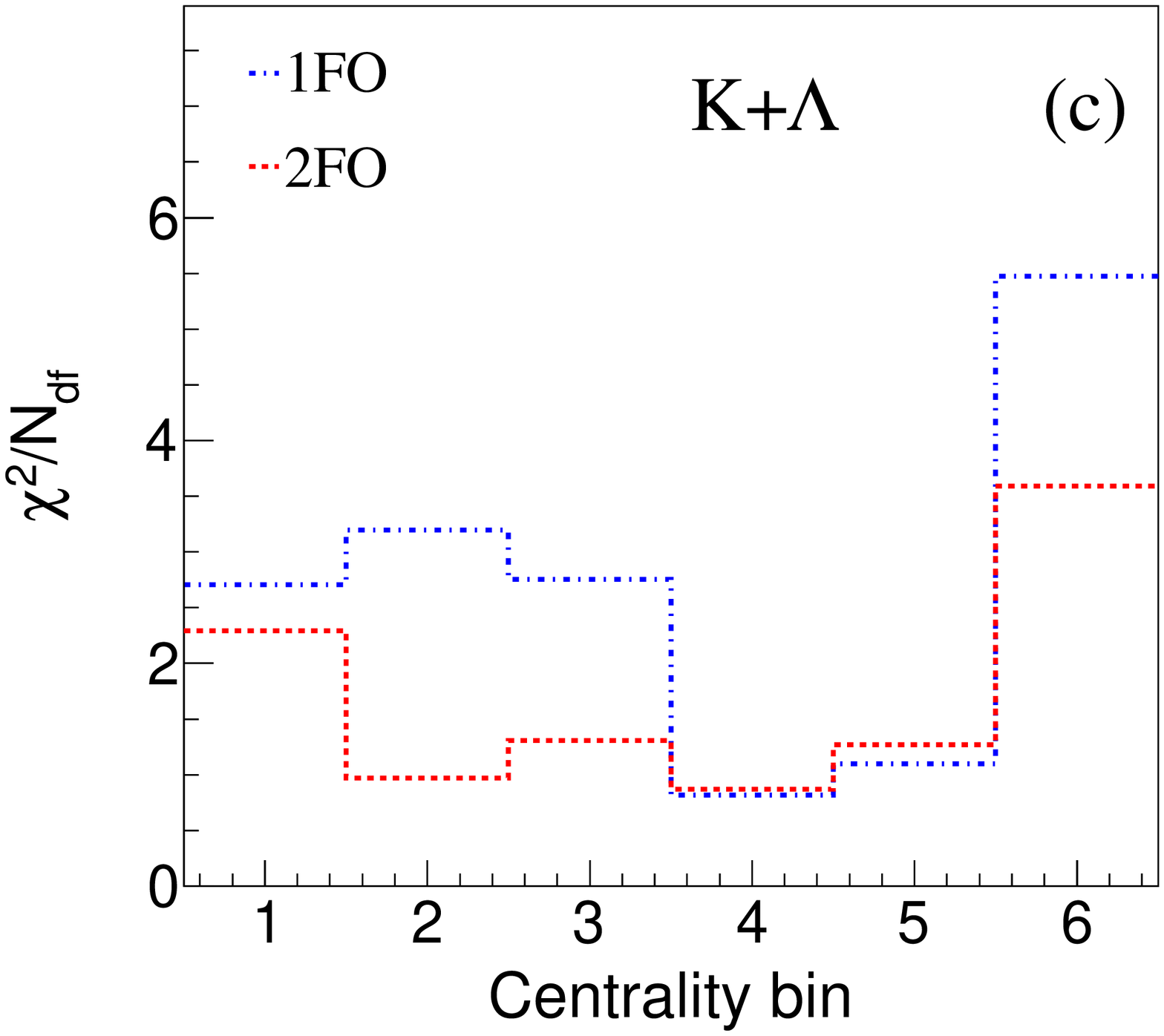}}    
  \caption{(Color online) Plots of $\chi^2/N_{df}$ in 1FO and 2FO. The left plot gives the total $\chi^2/N_{df}$ 
  of both FHs while the middle and right plots give the $\chi^2/N_{df}$ for the non-strange and strange FHs 
  respectively. The centrality bin numbers 1, 2, 3, 4, 5 and 6 refer to the $\l00-05\r\%$, $\l05-10\r\%$, 
  $\l10-20\r\%$, $\l20-40\r\%$, $\l40-60\r\%$ and $\l60-80\r\%$ centrality bins respectively.}
  \label{fig.chi2}
  \end{center}
  \end{figure}
  
  \begin{figure}[htb]
  \begin{center}
    \scalebox{0.29}{\includegraphics{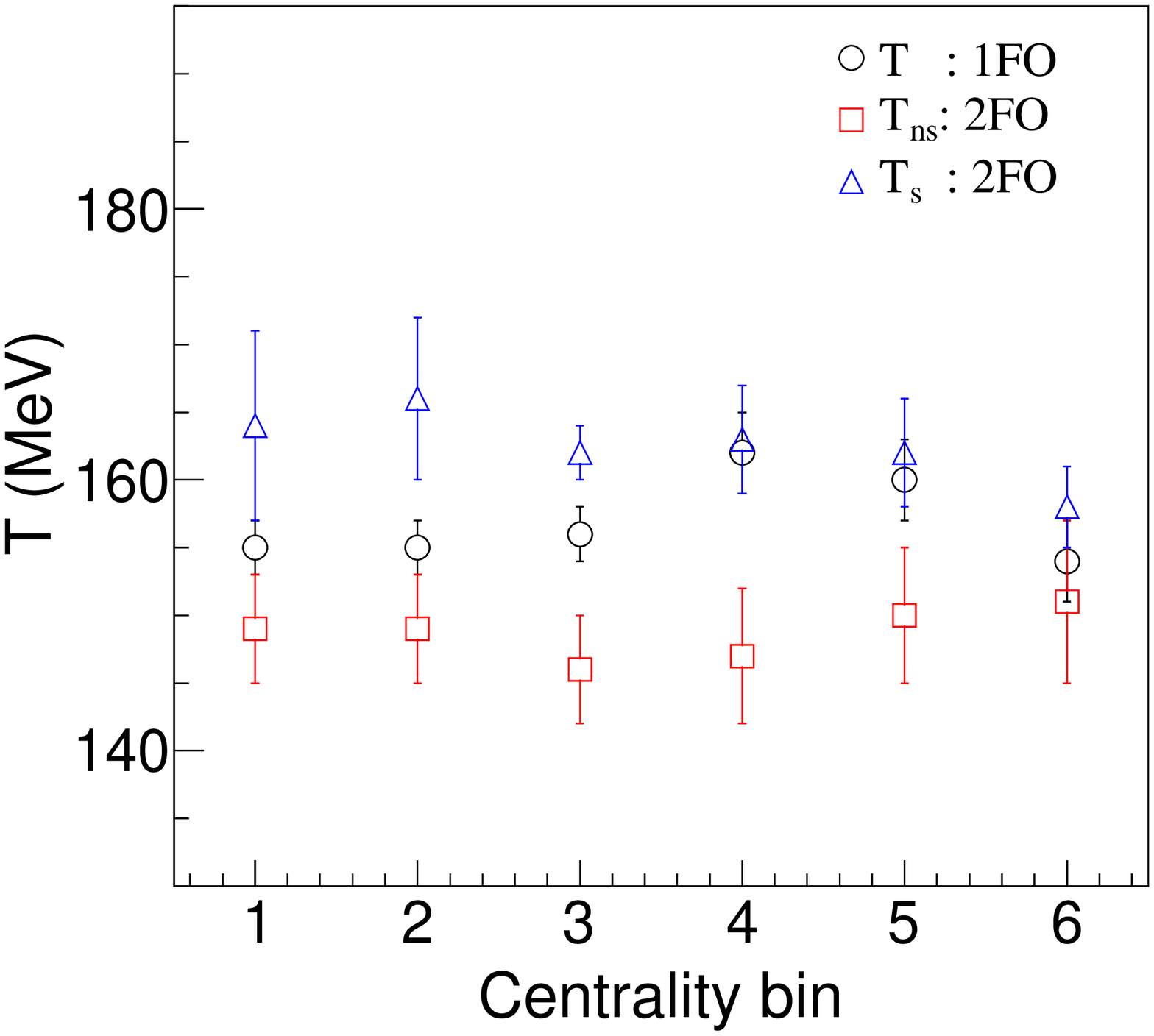}}
    \scalebox{0.29}{\includegraphics{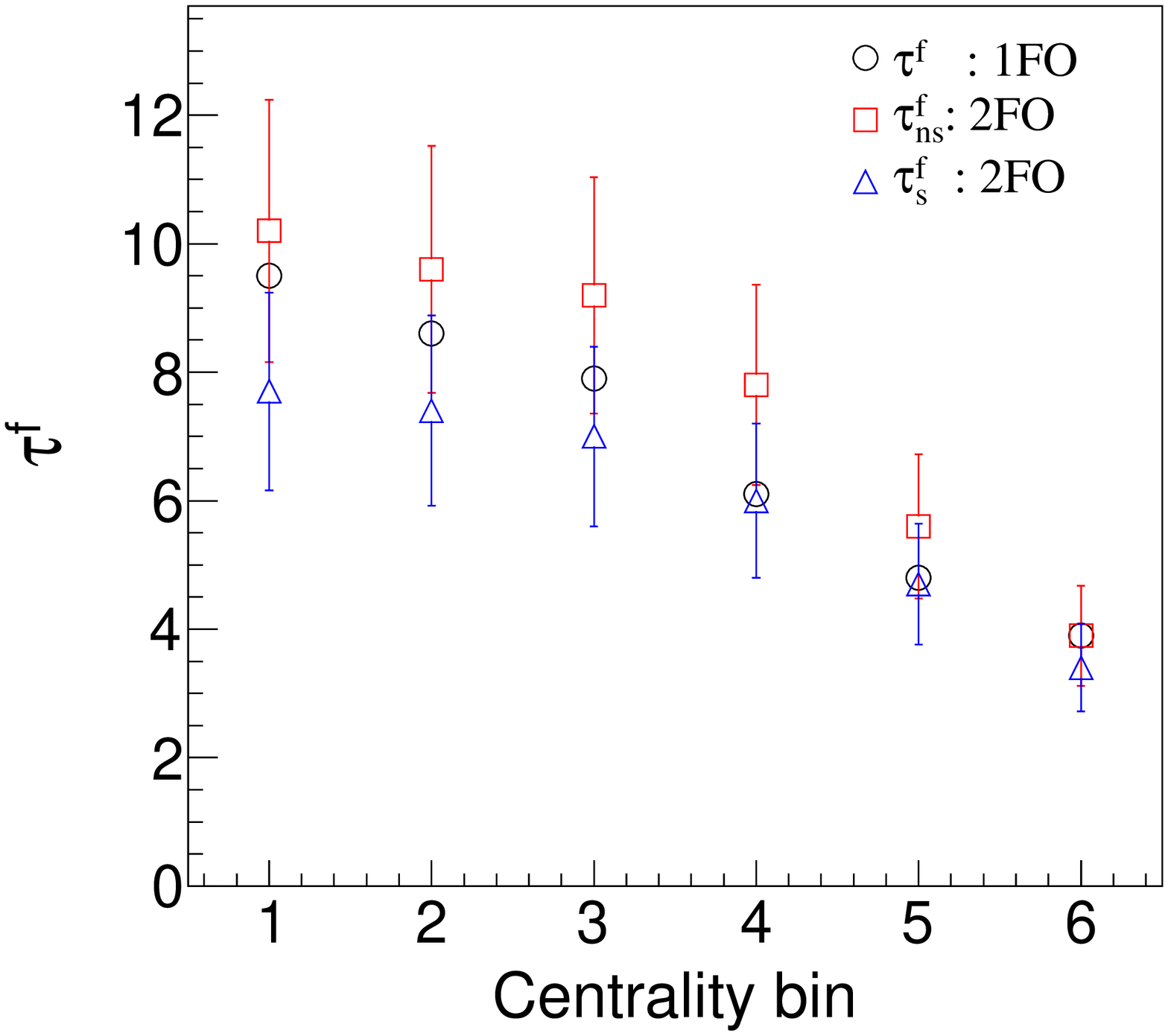}}
    \scalebox{0.29}{\includegraphics{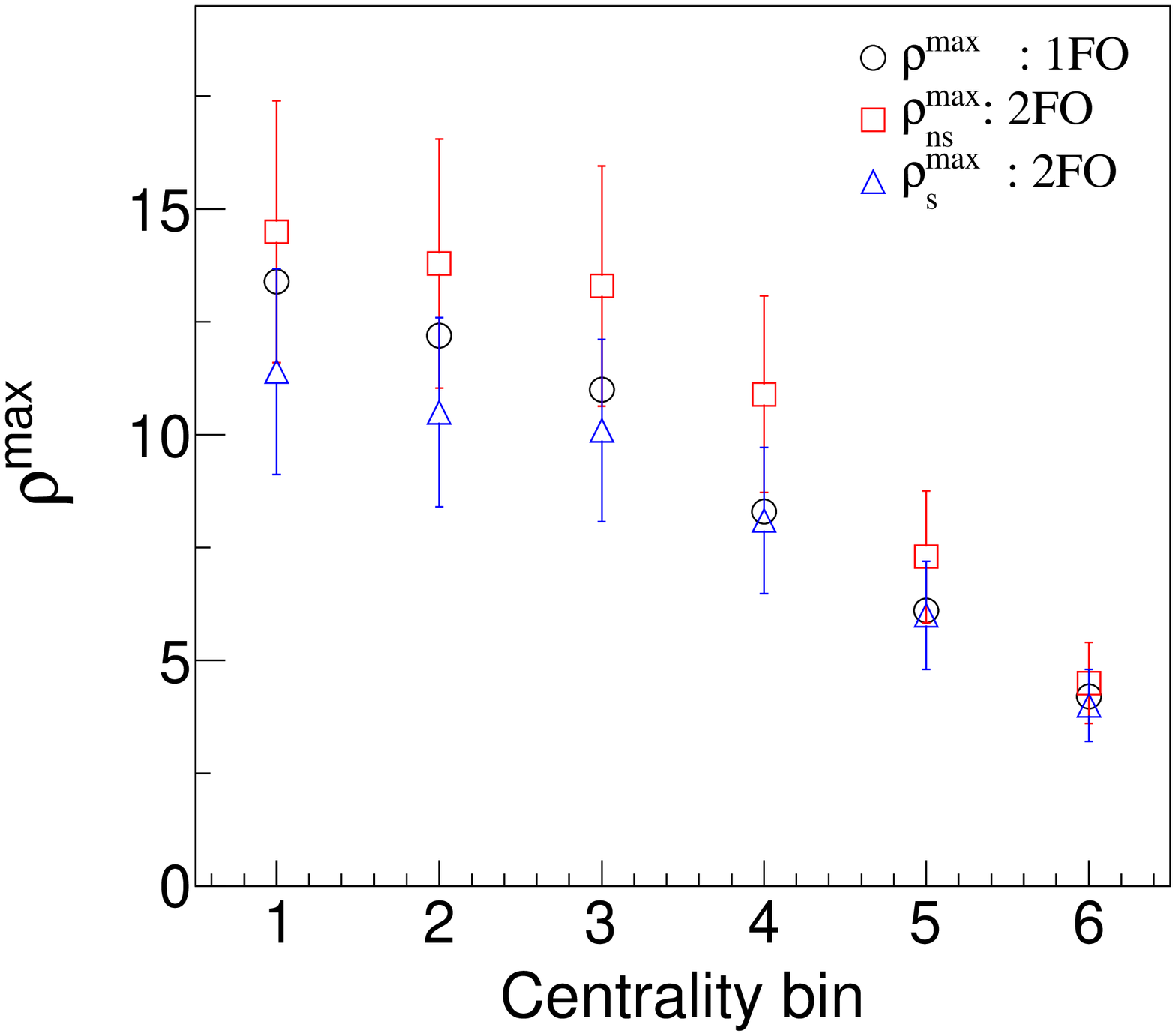}}    
  \caption{(Color online) Plots of the different freezeout parameters in 1FO and 2FO. In the left plot, $T$ is 
  plotted while the middle and right plots show the geometric parameters $\tauf$ and $\rhom$. The centrality bin 
  numbers 1, 2, 3, 4, 5 and 6 refer to the $\l00-05\r\%$, $\l05-10\r\%$, $\l10-20\r\%$, $\l20-40\r\%$, $\l40-60\r\%$ 
  and $\l60-80\r\%$ centrality bins respectively.}
  \label{fig.parameters}
  \end{center}
  \end{figure}

In Table~\ref{tab.2FOparams}, we have listed all the parameters that were extracted from fits to data on yields 
and spectra~\cite{Abelev:2013vea,Abelev:2013xaa,ABELEV:2013zaa,Abelev:2014uua}. The errors on the geometric 
parameters correspond to variation of $T$ within its errors. We first compare the goodness of fits for both 
the schemes in terms of $\chi^2/N_{df}$ where
\beqa
\chi^2&=&\sum_i\l\l \text{Data}\l {p_T}_i\r-\text{Model}\l {p_T}_i\r\r/\text{Error}\l {p_T}_i\r\r^2\nn\\
N_{df}&=&\text{Number of data points - Number of free parameters}\nn
\label{eq.chisq}
\eeqa
Since 1FO has three parameters while 2FO has six, we do not compare the $\chi^2$ in the two schemes directly.  
The reduced $\chi^2$ defined as $\chi^2/N_{df}$ is the appropriate statistical quantity to compare in this 
case as $N_{df}$ takes care of the difference in the number of free parameters.

We have performed the sum over all available data points upto $p_T = 3$ GeV/c. Overall, 2FO provides a better 
description of the $p_T$ spectra, particularly for $\pi^+,K^+,p$ and $\Lambda$. This is evident from the plot 
of $\chi^2$ vs $N_{df}$ in Fig.~\ref{fig.chi2}. In Fig.~\ref{fig.chi2} (a), $\chi^2/N_{df}$ summed up for the 
four hadrons $\pi^+,K^+,p$ and $\Lambda$ is shown. Both 1FO and 2FO schemes yield better fits for mid-central 
collisions compared to more central and peripheral collisions. For e.g. in 2FO, centrality bin numbers 3, 4 
and 5 have $\chi^2/N_{df}<2$ while for the rest, $\chi^2/N_{df}$ ranges between 2 to 4. Similar conclusion was 
also drawn in a study within the NFO scheme~\cite{Begun:2014rsa}. The best fit is for the $\l40-60\r\%$ 
centrality bin with $\chi^2/N_{df}~1.5$. 
In all the centrality bins, there is a reduction in $\chi^2/N_{df}$ for 2FO as compared to 1FO. The maximum 
decrease by around 2.5 units is for the most central bin. For the next three bins, there is a similar reduction 
of 1.5 units. For the $\l40-60\r\%$ centrality bin, the decrease is by only around 0.7 unit. Finally, for the 
most peripheral bin, the reduction is by around one unit. It is interesting to analyse the split up of the 
$\chi^2$ between the non-strange and strange FHs. These are shown in Figs.~\ref{fig.chi2} (b) and 
\ref{fig.chi2} (c). Except for the most peripheral bin, in case of both 1FO and 2FO, the $\chi^2/N_{df}$ of the 
strange FH is much lower than that of the non-strange FH.

The fitted parameters in 1FO and 2FO are shown in Fig.~\ref{fig.parameters}. For all centralities, the 1FO 
parameters are flanked by the 2FO parameters. While for the more central bins like $\l00-05\r \%$, $\l05-10\r \%$ 
and $\l10-20\r \%$, the 1FO parameter is close to the mean of the strange and non-strange parameters, for the 
remaining bins the 1FO parameters lie close to the strange FH parameters. This trend is there in the $T$ extracted 
from the yields and gets rubbed off into the geometric parameters as well. The strange and non-strange parameters 
seem to come close to each other as one goes from central to peripheral collisions, implying sudden and simultaneous 
freezeouts of the fireball for the peripheral collisions in comparison to a more gradual and 
sequential freezeout for the central collisions. We also note that while the temperatures of the strange and non-
strange FHs are almost constant within errors throughout all the centralities, the geometric parameters monotonically 
decrease from central to peripheral collisions. This implies that while the temperature of the last scattering surface 
is similar, the fireball shrinks in volume as one goes from central to peripheral collisions. We will now focus on 
the spectra of individual hadrons.

  \begin{figure}[htb]
  \begin{center}
    \scalebox{0.5}{\includegraphics{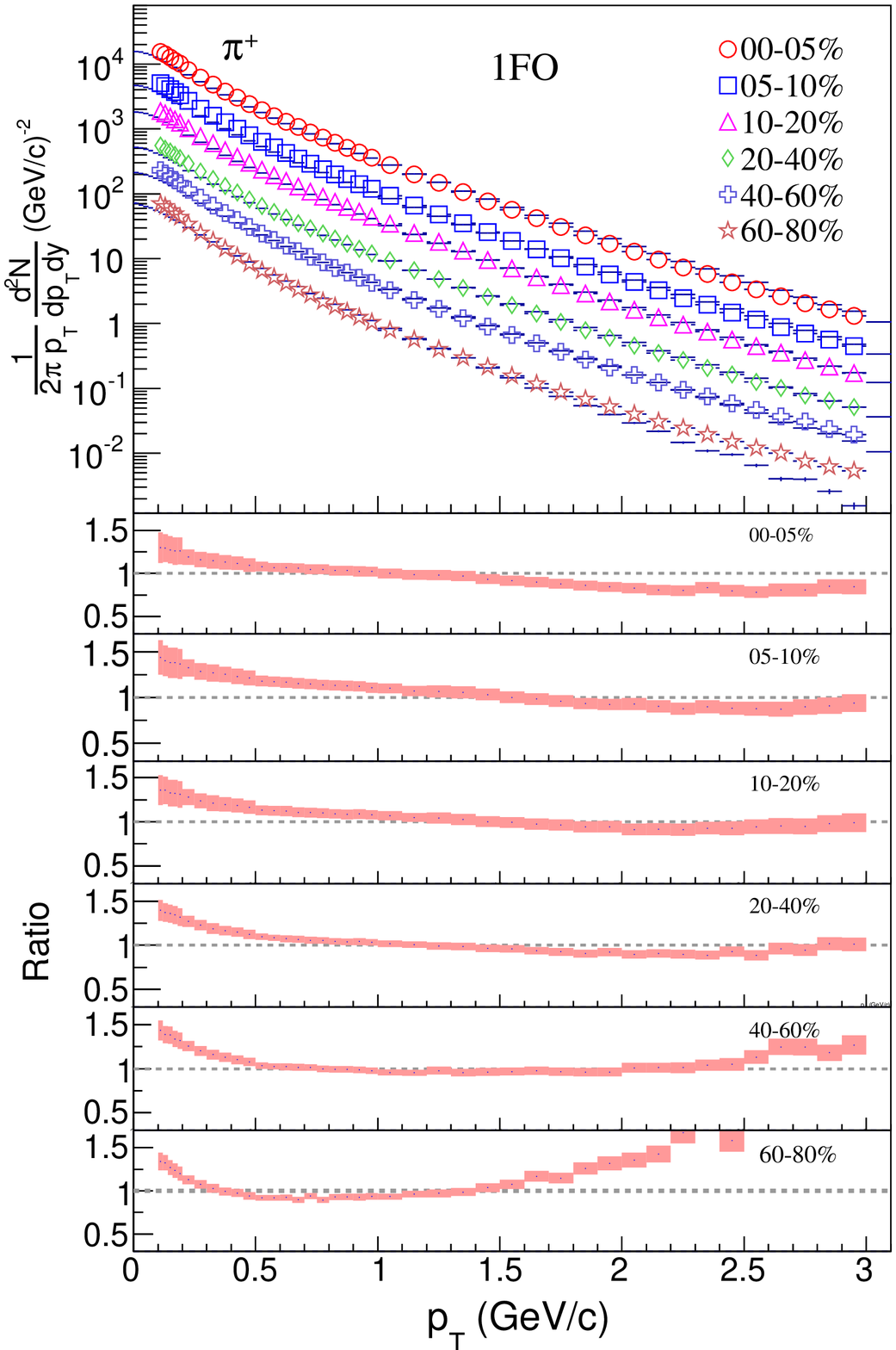}}
    \scalebox{0.5}{\includegraphics{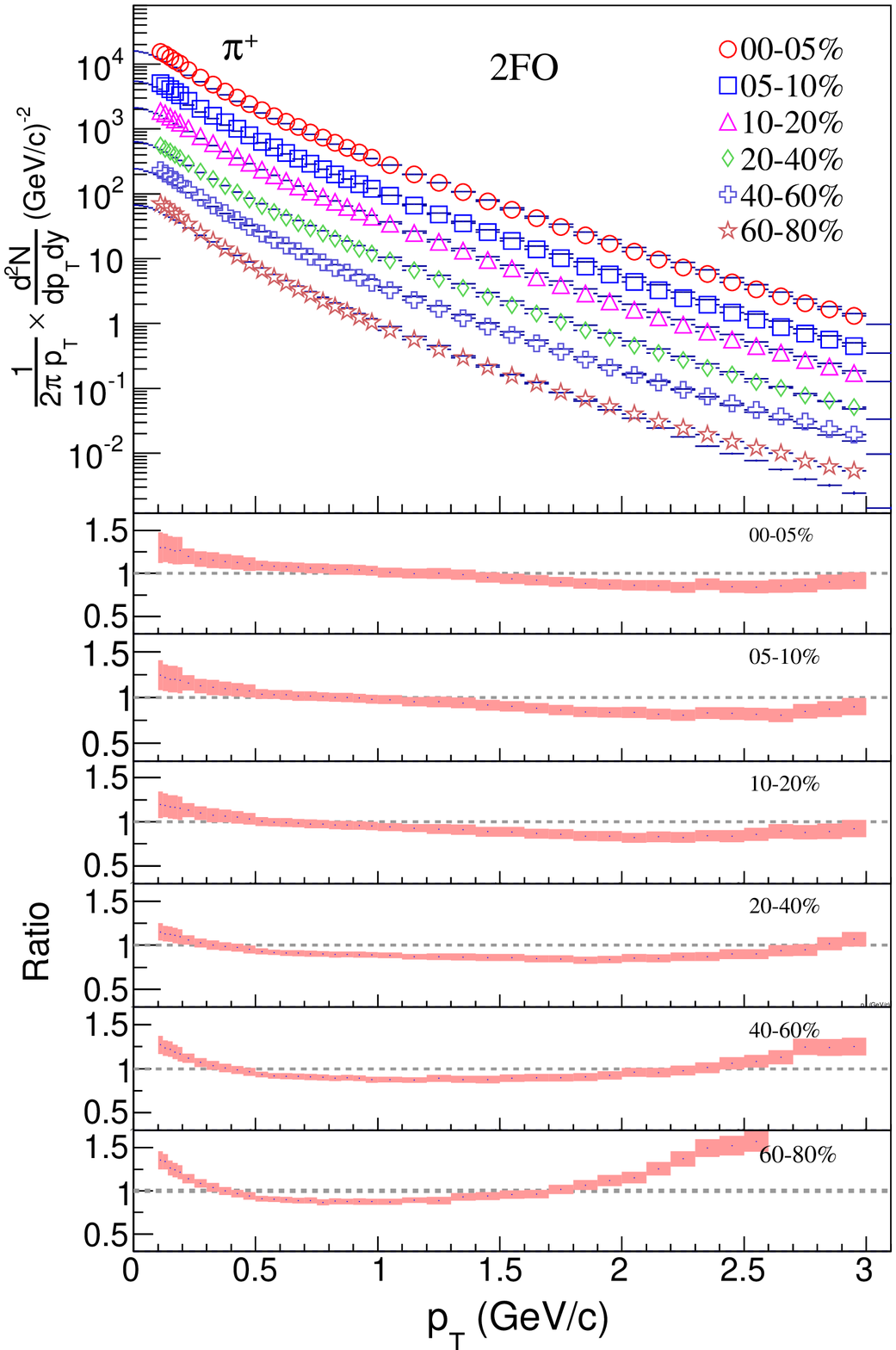}}  
  \caption{(Color online) Plots of $\pi^+$ spectra as obtained in 1FO (left) vs 2FO (right). While the top panels 
  give the comparison of the $p_T$ spectra between data~\cite{Abelev:2013vea} and model, the lower panels plot 
  the ratio = $Data/Model$ for the different centrality bins.
}
  \label{fig.pi}
  \end{center}
  \end{figure}
  
   \begin{figure}[htb]
  \begin{center}
    \scalebox{0.5}{\includegraphics{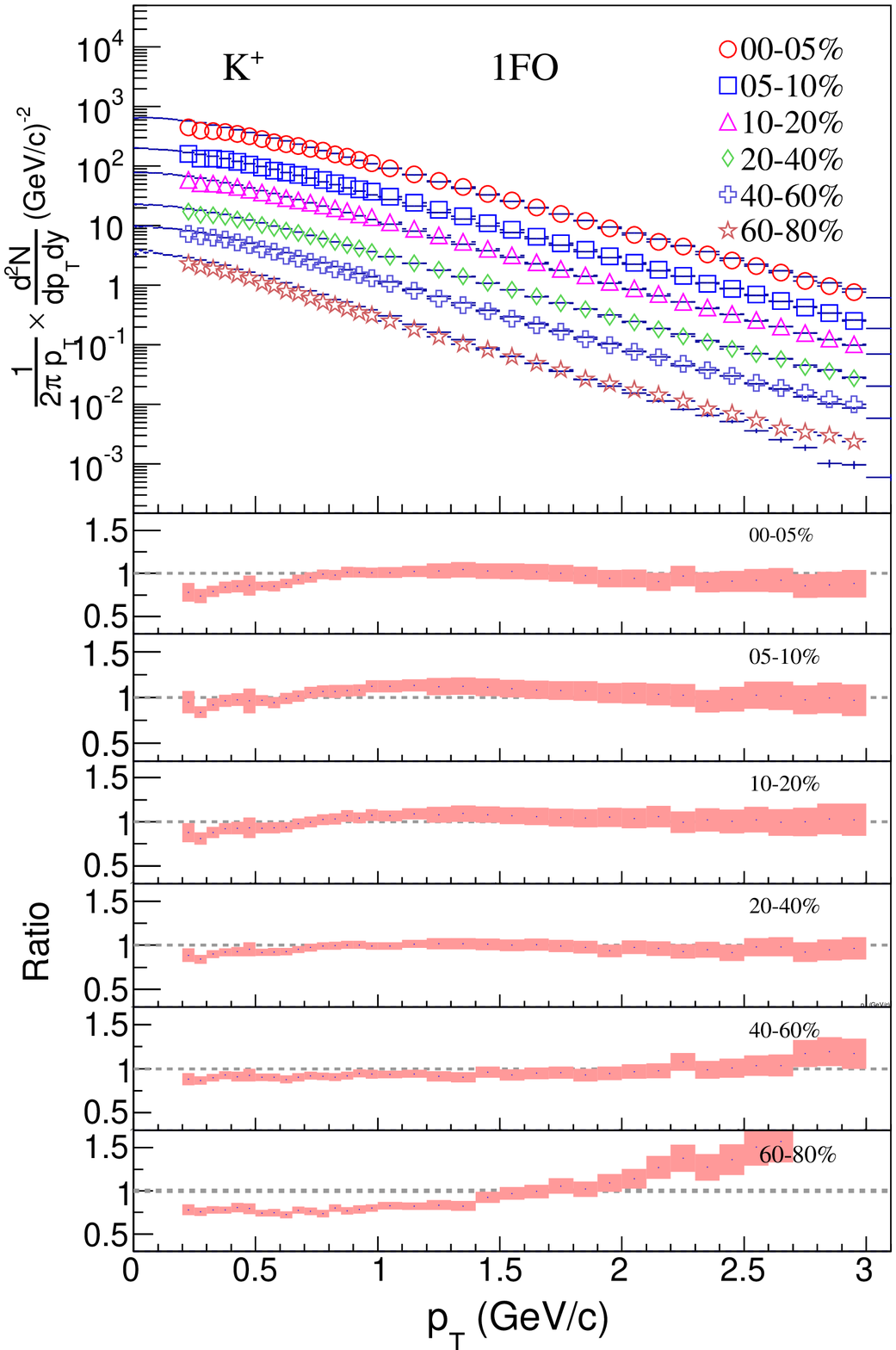}}
    \scalebox{0.5}{\includegraphics{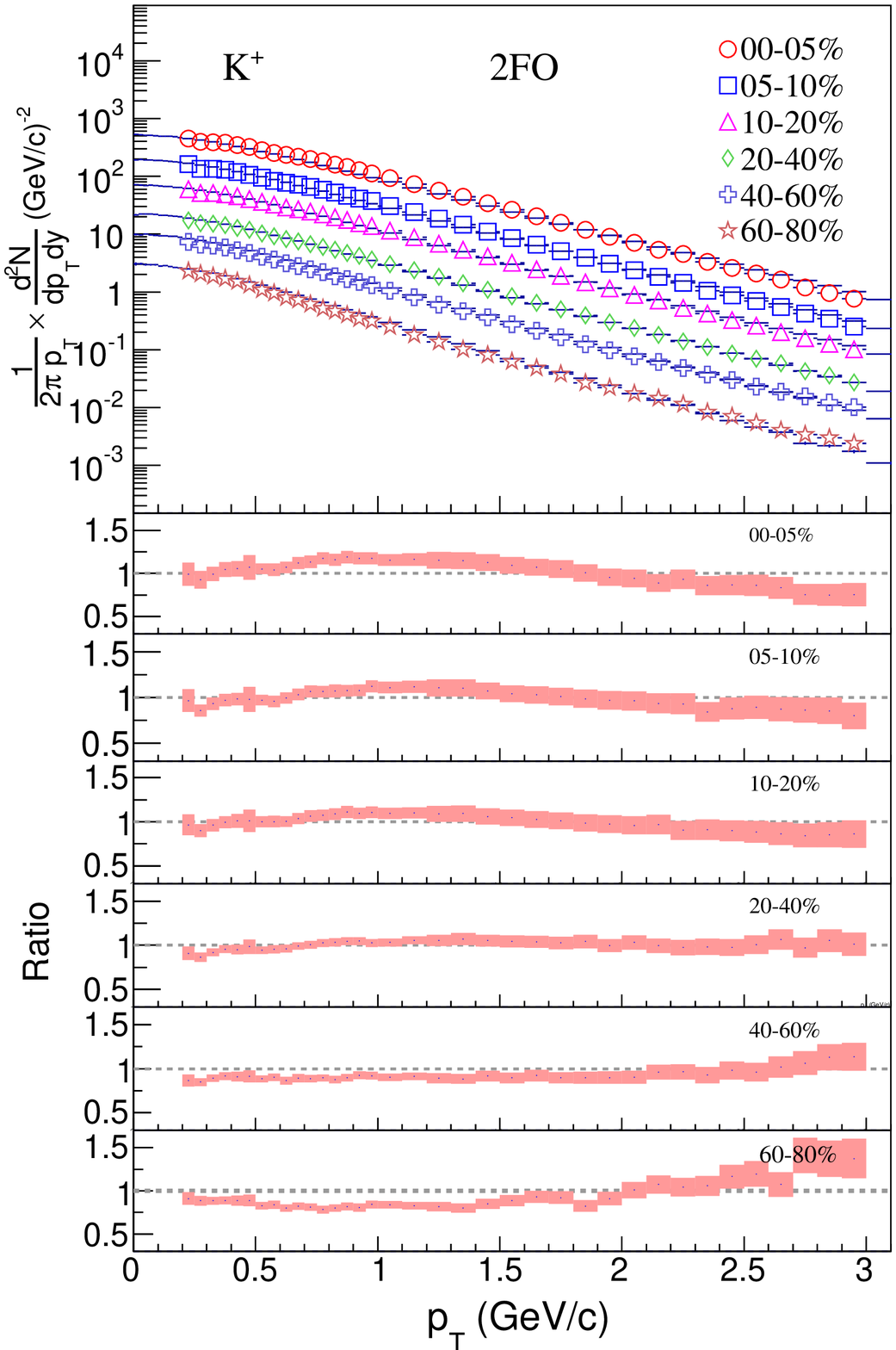}}  
  \caption{(Color online) Plots of $K^+$ spectra as obtained in 1FO (left) vs 2FO (right). While the top panels 
  give the comparison of the $p_T$ spectra between data~\cite{Abelev:2013vea} and model, the lower panels plot 
  the ratio = $Data/Model$ for the different centrality bins.
}
  \label{fig.kaon}
  \end{center}
  \end{figure}
  
    \begin{figure}[htb]
  \begin{center}
    \scalebox{0.5}{\includegraphics{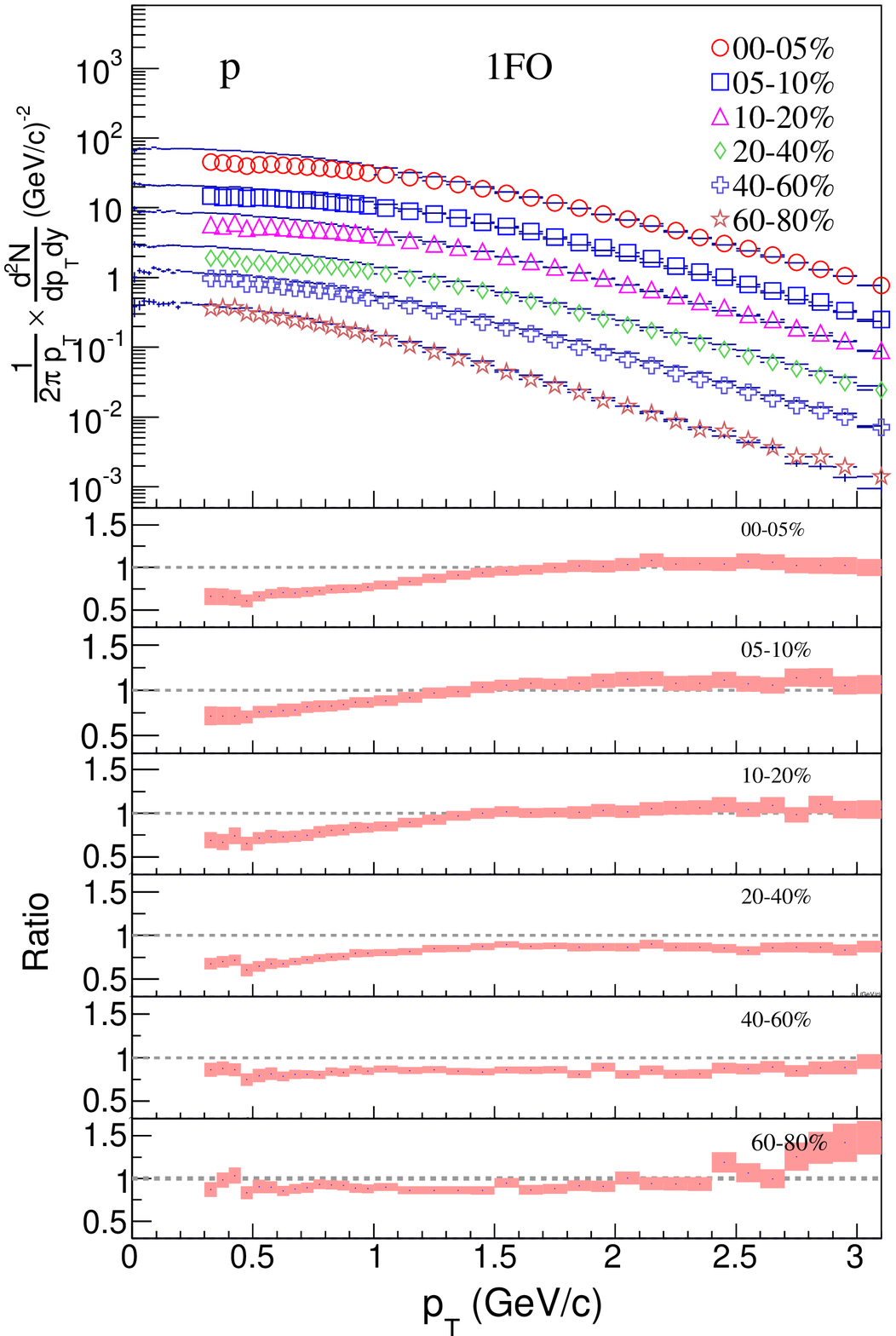}}
    \scalebox{0.5}{\includegraphics{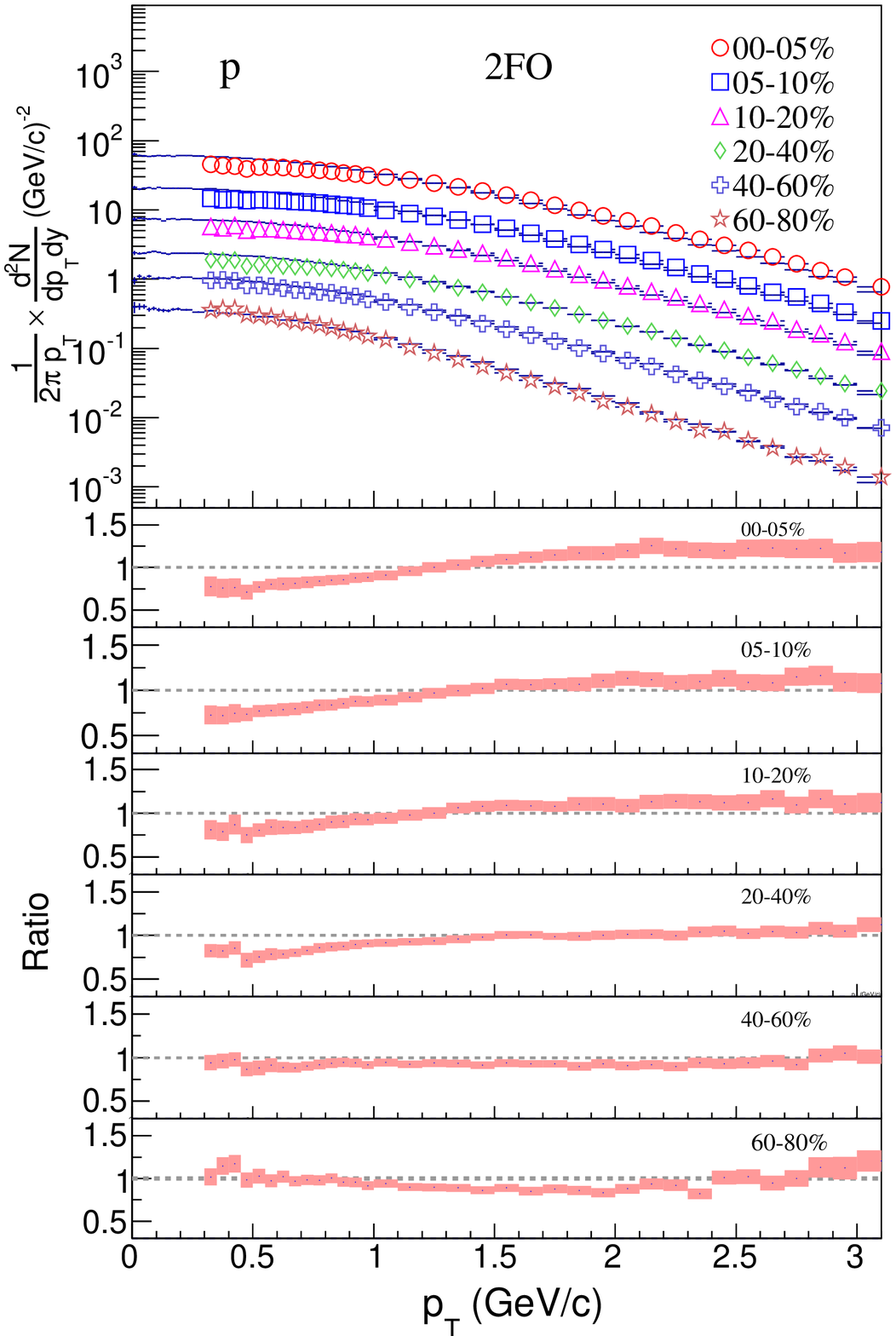}}  
  \caption{(Color online) Plots of $p$ spectra as obtained in 1FO (left) vs 2FO (right). While the top panels 
  give the comparison of the $p_T$ spectra between data~\cite{Abelev:2013vea} and model, the lower panels plot 
  the ratio = $Data/Model$ for the different centrality bins.
}
  \label{fig.proton}
  \end{center}
  \end{figure}
  
    \begin{figure}[htb]
  \begin{center}
    \scalebox{0.5}{\includegraphics{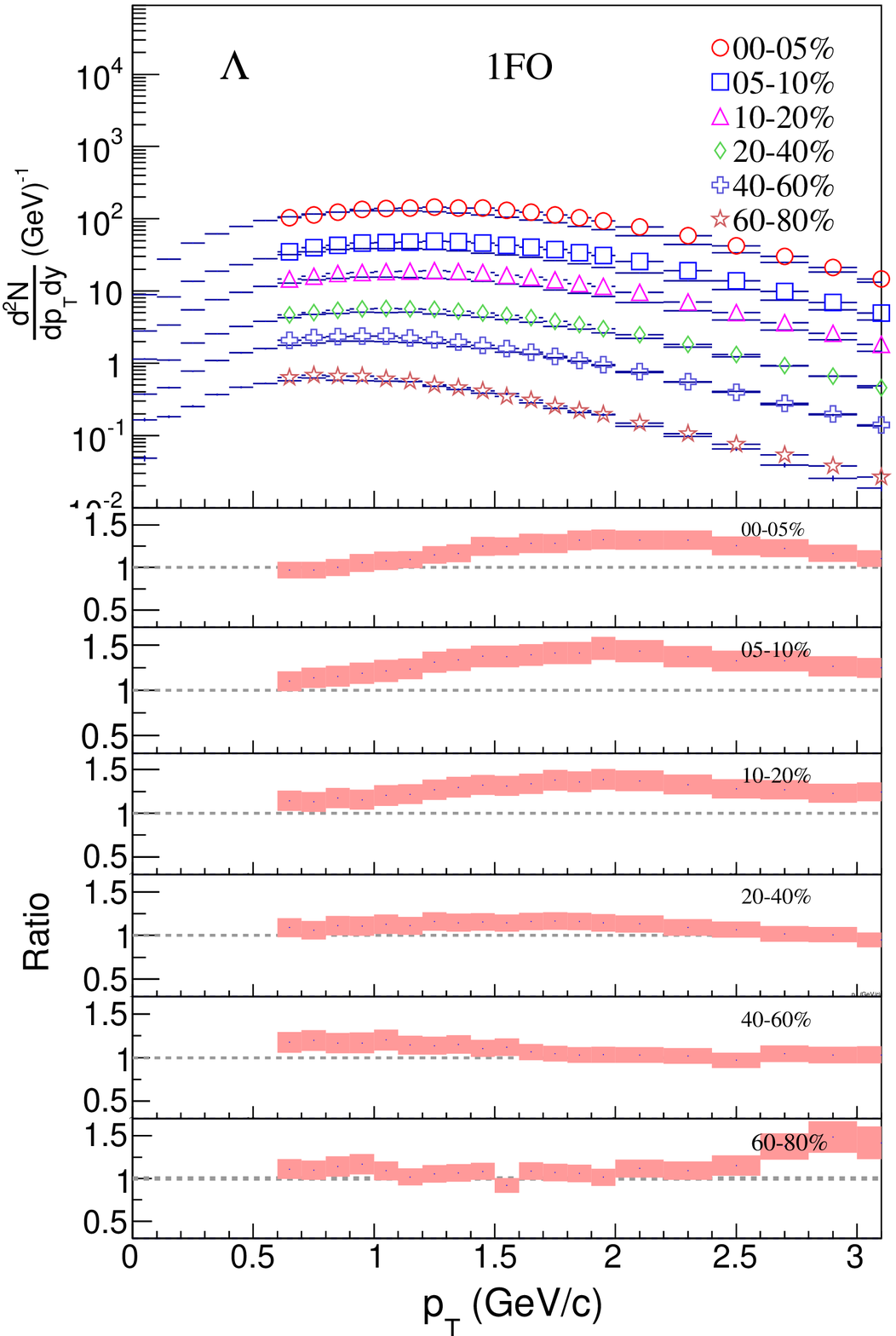}}
    \scalebox{0.5}{\includegraphics{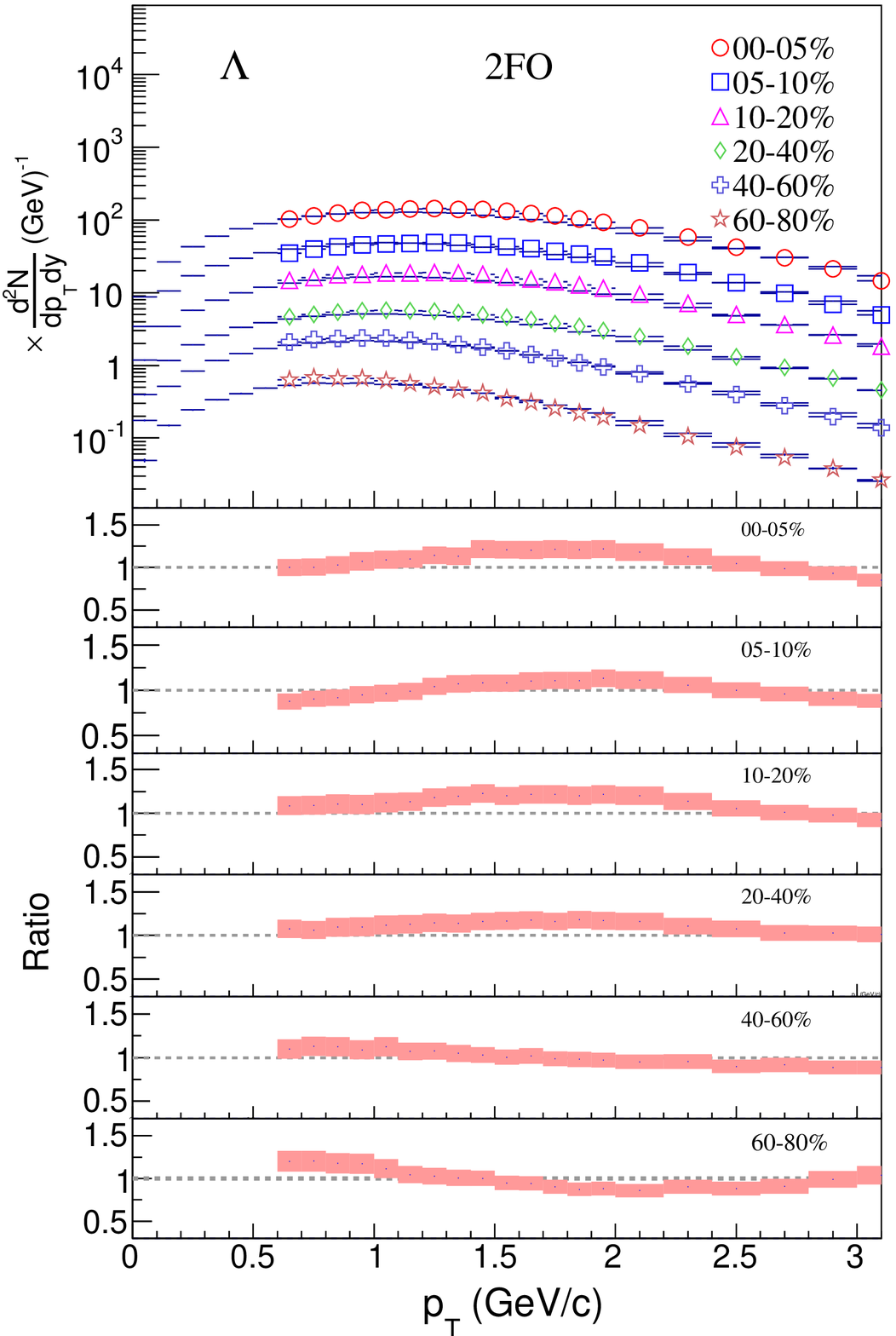}}  
  \caption{(Color online) Plots of $\Lambda$ spectra as obtained in 1FO (left) vs 2FO (right). While the top panels 
  give the comparison of the $p_T$ spectra between data~\cite{Abelev:2013xaa} and model, the lower panels plot 
  the ratio = $Data/Model$ for the different centrality bins.
}
  \label{fig.lambda}
  \end{center}
  \end{figure}
  
    \begin{figure}[htb]
  \begin{center}
    \scalebox{0.4}{\includegraphics{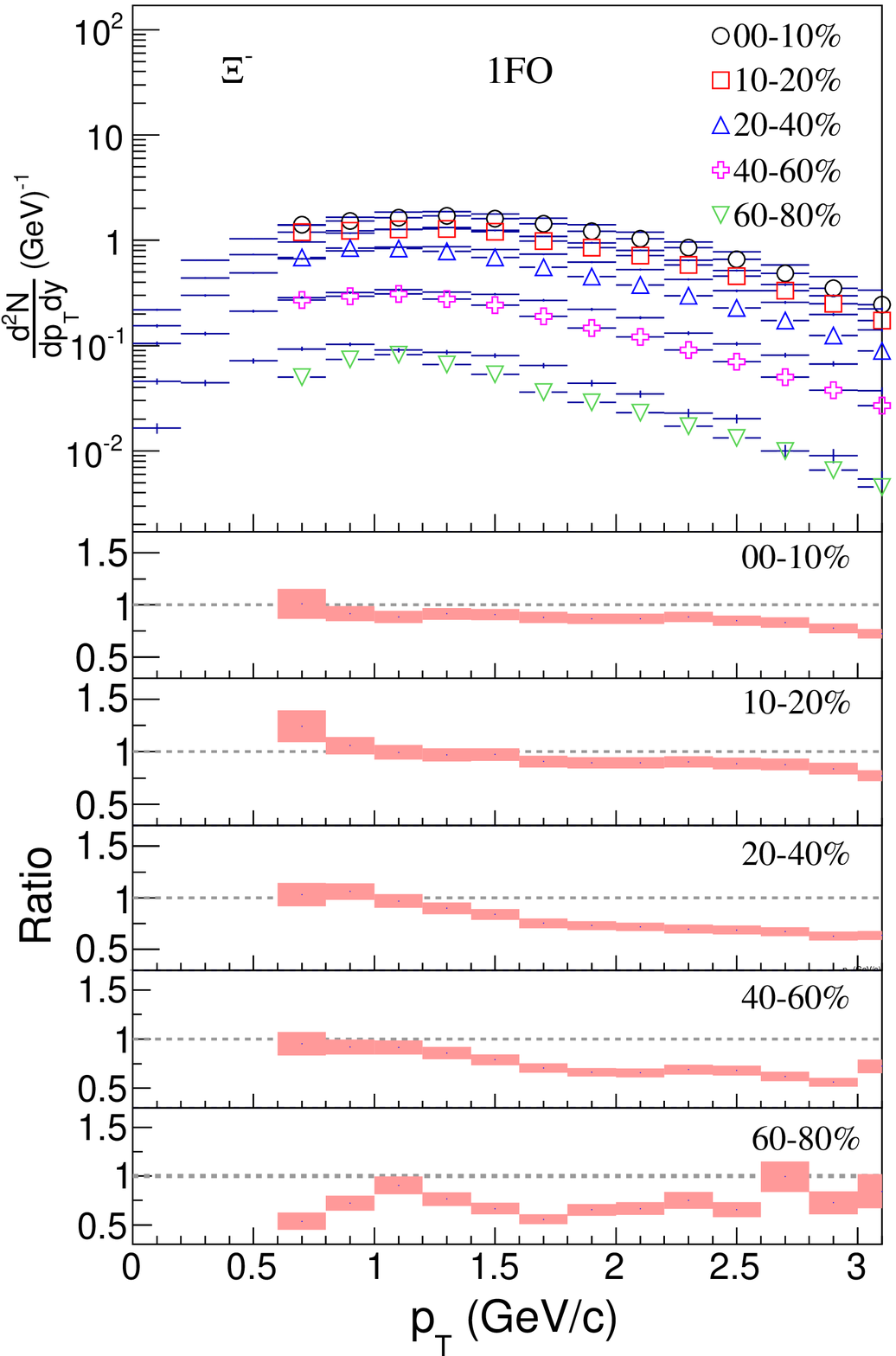}}
    \scalebox{0.4}{\includegraphics{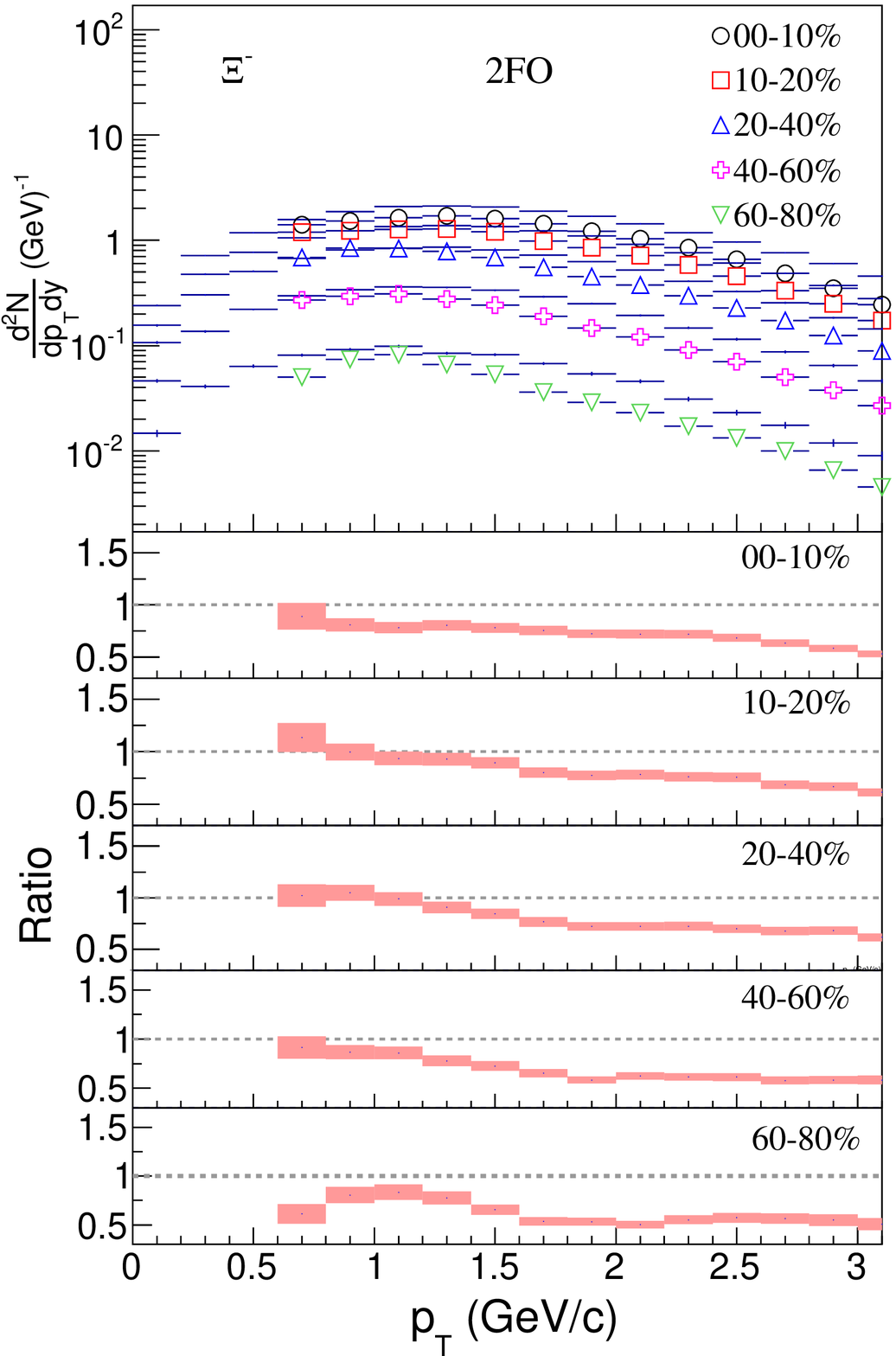}}  
    \scalebox{0.4}{\includegraphics{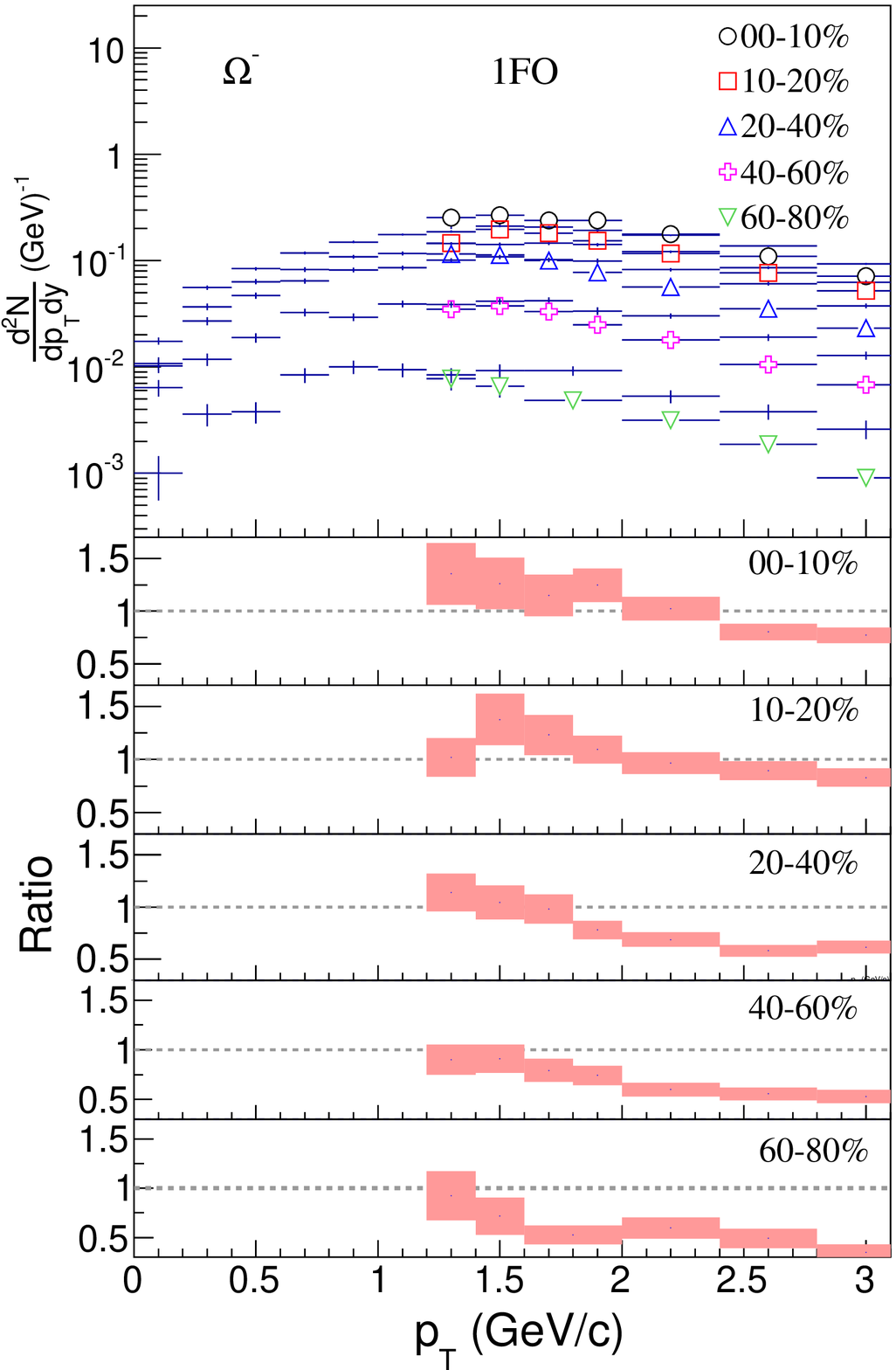}}
    \scalebox{0.4}{\includegraphics{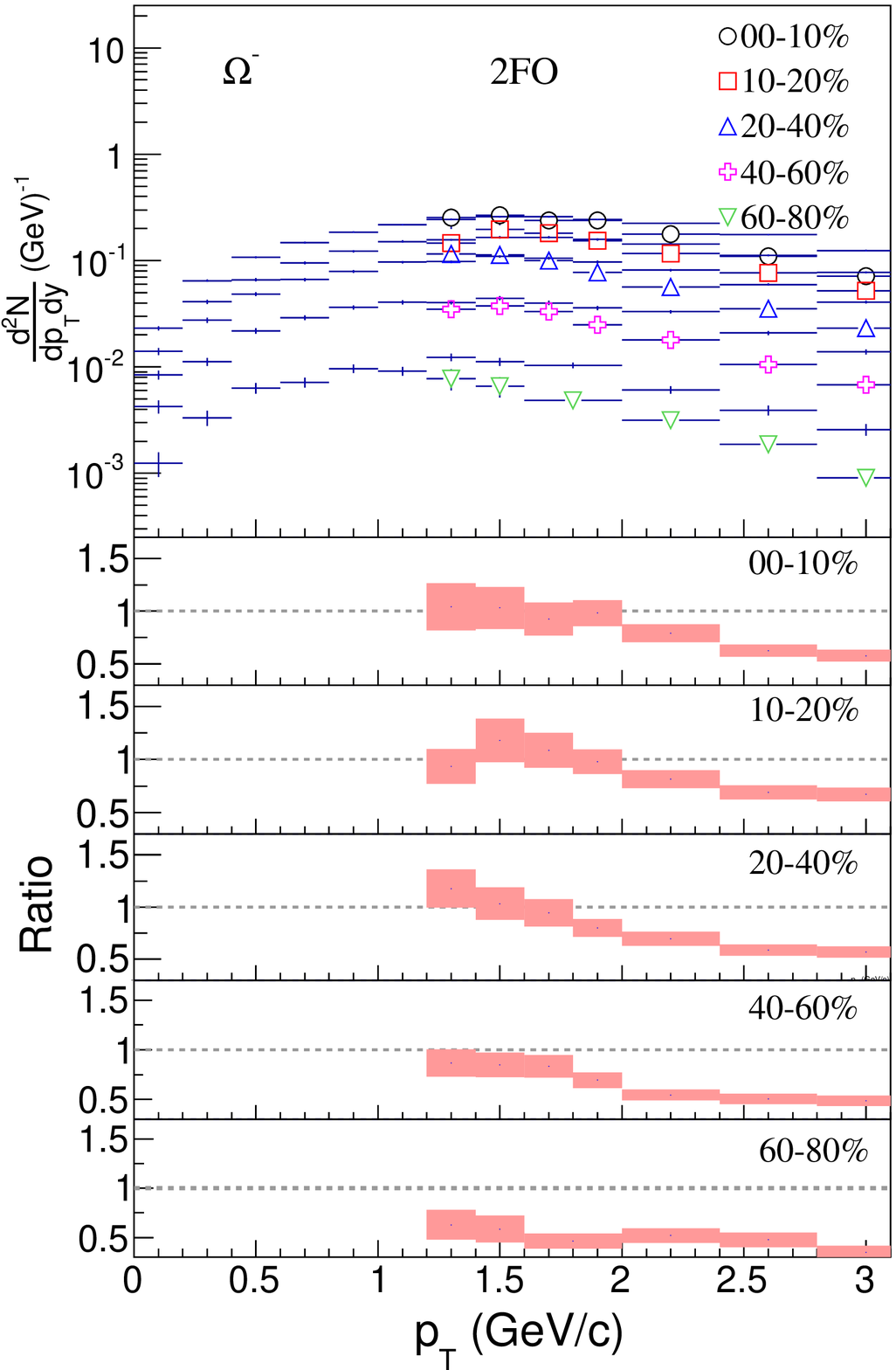}} 
  \caption{(Color online) Plots of $\Xi^-$ (upper) and $\Omega$ (lower) spectra as obtained in 1FO (left) 
  vs 2FO (right). In each plot, the top panels give the comparison of the $p_T$ spectra between data~\cite{ABELEV:2013zaa} 
  and model and the lower panels plot the ratio = $Data/Model$ for the different centrality bins.
}
  \label{fig.multistrange}
  \end{center}
  \end{figure}

    \begin{figure}[htb]
  \begin{center}
    \scalebox{0.43}{\includegraphics{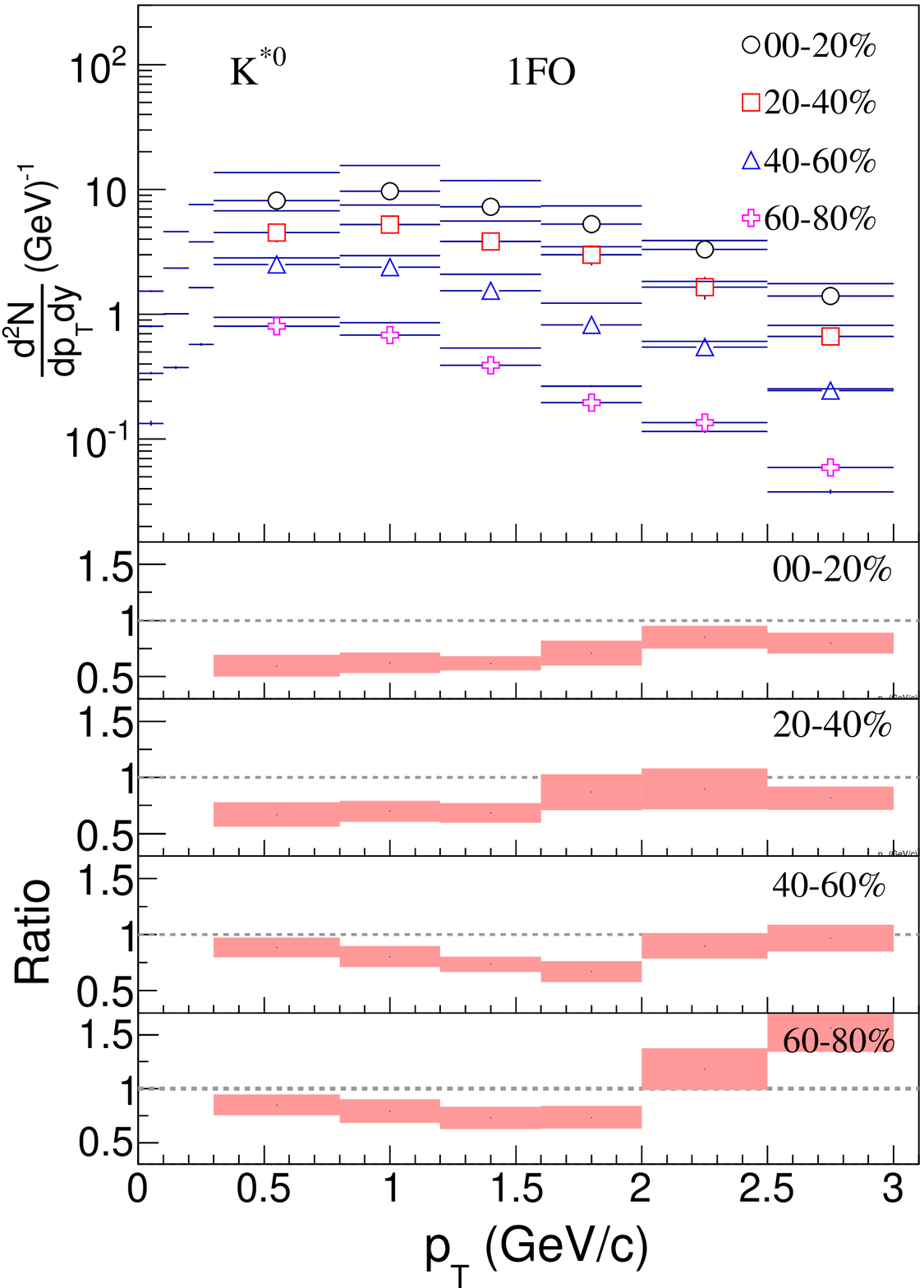}}
    \scalebox{0.43}{\includegraphics{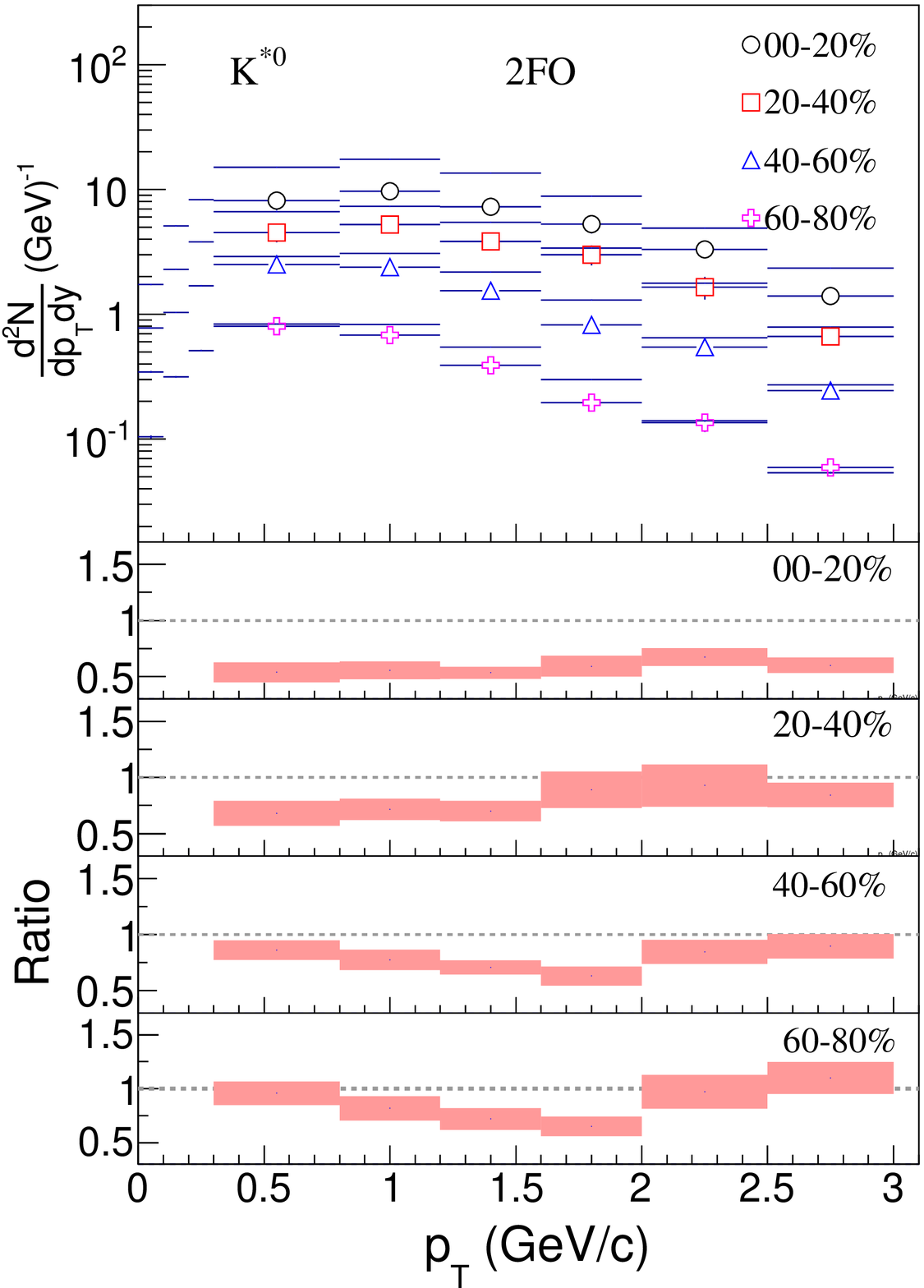}}  
  \caption{(Color online) Plots of $K^*$ spectra as obtained in 1FO (left) vs 2FO (right). While the top panels 
  give the comparison of the $p_T$ spectra between data~\cite{Abelev:2014uua} and model, the lower panels plot 
  the ratio = $Data/Model$ for the different centrality bins.
}
  \label{fig.kstar}
  \end{center}
  \end{figure}

   \begin{figure}[htb]
  \begin{center}
    \scalebox{0.5}{\includegraphics{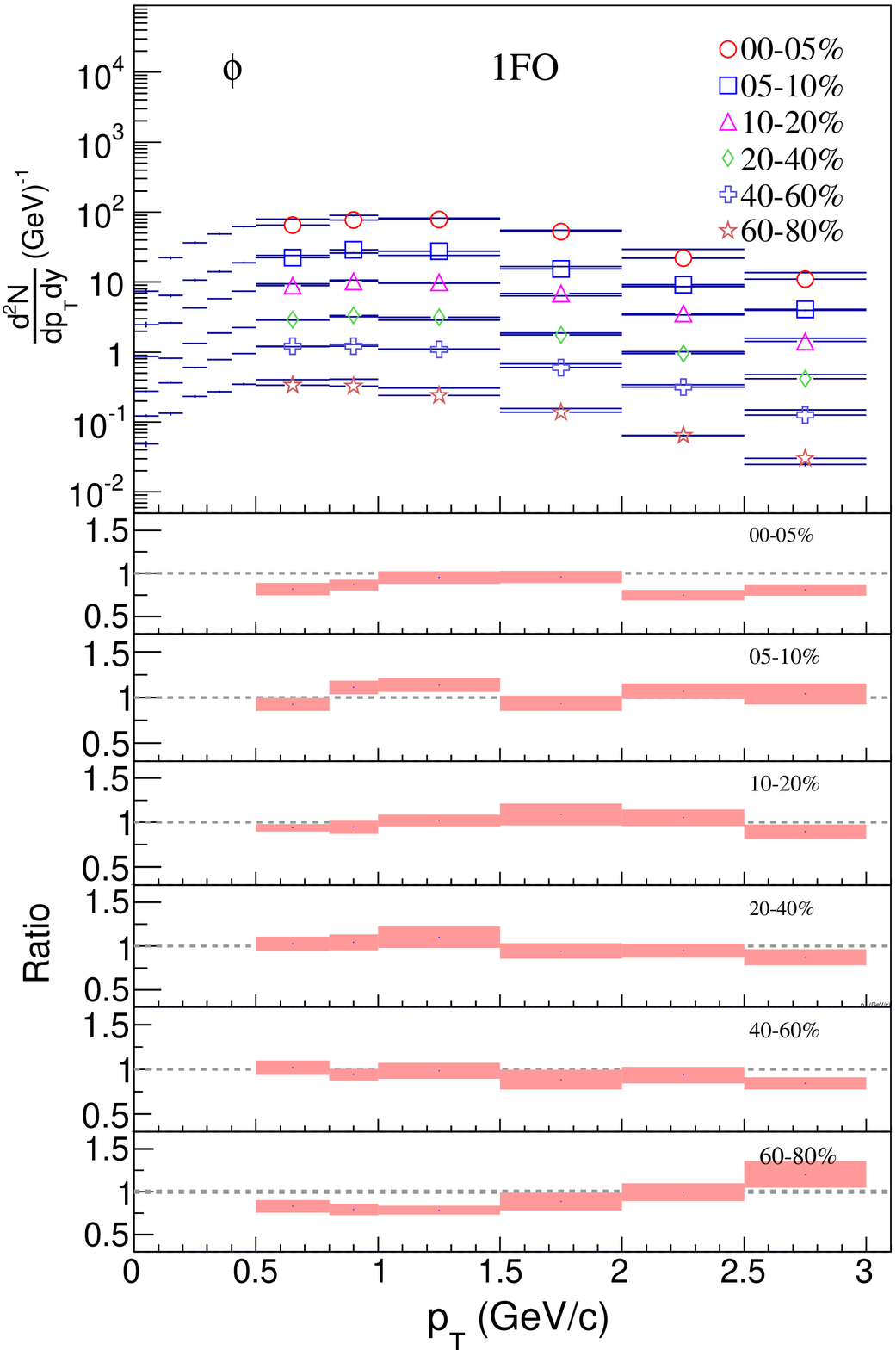}}
    \scalebox{0.5}{\includegraphics{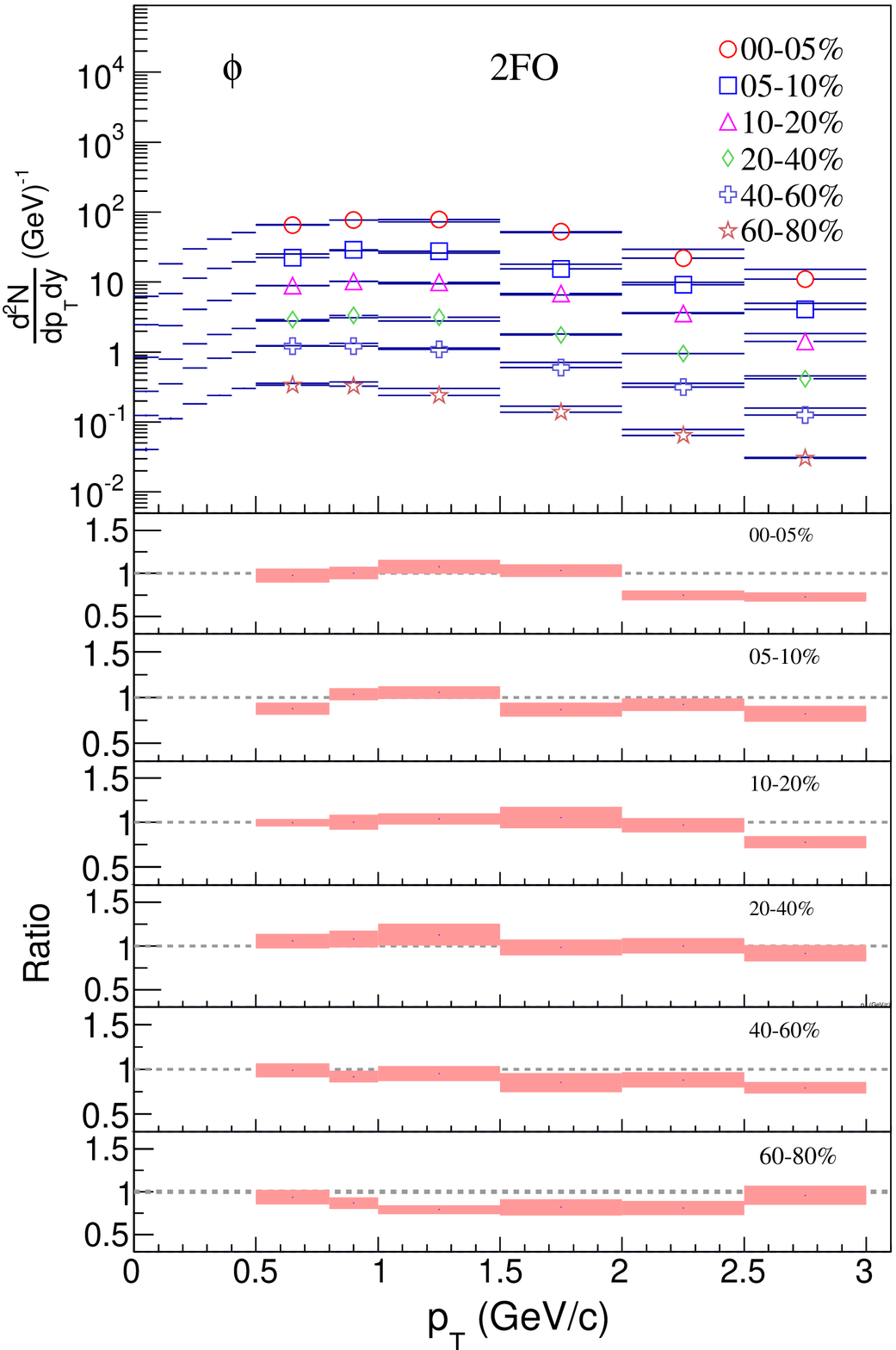}}  
  \caption{(Color online) Plots of $\phi$ spectra as obtained in 1FO (left) vs 2FO (right). While the top panels 
  give the comparison of the $p_T$ spectra between data~\cite{Abelev:2014uua} and model, the lower panels plot 
  the ratio = $Data/Model$ for the different centrality bins.
}
  \label{fig.phi}
  \end{center}
  \end{figure}

The model $p_T$ spectra in 1FO and 2FO and their comparison with data are shown for the cases of $\pi^+$, $K^+$, 
$p$, $\Lambda$, $\Xi^-$, $\Omega$, $K^*$ and $\phi$ in Figs.~\ref{fig.pi}, \ref{fig.kaon}, \ref{fig.proton}, 
\ref{fig.lambda}, \ref{fig.multistrange}, \ref{fig.kstar} and \ref{fig.phi}.
In 2FO, we find that the agreement between data and model for all the centralities with $p_T$ range upto 2 GeV/c is
within $25\%$ for all the hadrons except the multi-strange baryons and $K^*$ resonance. In Figs.~\ref{fig.pi} and 
\ref{fig.proton}, we see that in 1FO, there is a strong tension between the model spectra of $\pi^+$ and $p$ for 
low and high $p_T$ : in low (high) $p_T$ the model underpredicts (overpredicts) in case of $\pi^+$ while for $p$, 
the behaviour is just the reverse of $\pi^+$ . In 2FO, this tension is reduced considerably. This is the main source 
of improvement in the fits between 1FO and 2FO. However, the tension does not go away completely in 2FO. There is a 
clear hint that the model still underpredicts at low $p_T$, particularly for the most central and peripheral 
collisions. A similar exercise within the NFO scheme gives a better description of 
the low $p_T$ pions~\cite{Begun:2014rsa}. It is interesting to note that the value of the light quark nonequilibrium 
phase space parameter obtained from fit to yield data results in pion chemical potential which is close to pion mass. 
This could suggest formation of pion condensation~\cite{Begun:2014rsa}. However, within the equilibrium chemical freezeout 
scenario, this could be interpreted as an indication for further substructure in freezeout and $\pi^+$ and $p$ freezeout 
separately. There could be also an alternate explanation. The decay of heavier resonances mainly contribute to 
low $p_T$ $\pi^+$. Thus, a deficit in low $p_T$ $\pi^+$ could also mean that currently the model does not include all 
the heavier resonances, hinting at the presence of yet unobserved resonances~\cite{Bazavov:2014xya}. The spectra 
of $K^+$ and $\Lambda$ which freezeout together in 2FO are shown in Figs.~\ref{fig.kaon} and \ref{fig.lambda}. 
The agreement between model and data is better compared to $\pi^+$ and $p$. This is also clear from 
Fig.~\ref{fig.chi2} (c) where we see that in 2FO, except for the centrality bin numbers 1 and 6, for all the 
other centralities, $\chi^2/N_{df}\sim1$. The spectra of the multi-strange baryons is shown in 
Fig.~\ref{fig.multistrange}. In case of the $\Xi^-$ ($\Omega$), the agreement between data and model for 
$p_T$ upto 1.5 (2.0) GeV/c is within $25\%$ for all centralities except the most peripheral 
one. Moreover, 1FO seems to give a better description in the first two centrality bins over the entire $p_T$ range. 
Within the multiple freezeout scenario, better model description of the spectra plots of the multi-strange hadrons is 
possible by allowing them to undergo an earlier freezeout compared to $K$ and $\Lambda$. Here we would like to note 
that the NFO scheme fails to describe the spectra of all strange baryons. It is only after 
invoking an early freezeout of the strange baryons that the NFO scheme manages to describe the 
$p_T$ spectra of the strange baryons~\cite{Begun:2014rsa}. However, it is important to note that within the 
NFO picture, even after the freezeout of the strange baryons $p$ and $K$ could still 
interact and produce $\Lambda$. It would be interesting to check whether including the effects of yet unobserved 
strange hadrons which are predicted by quark models and indicated in lattice QCD simulations improve the agreement 
between data and model of the spectra of the multi-strange baryons~\cite{Bazavov:2014xya}.

In case of the resonances, both 1FO and 2FO provide good description of the spectra of $\phi$ which has zero net 
strangeness while for $K^*$ they fail. In the latter case the lifetime of $K^*$ which is about 4 fm is of the same 
order as that of the fireball lifetime and hence effects like rescattering and regeneration modify the yield and 
also the spectra considerably. Previous comparison of the $K^*$ yield between models based on thermal production 
and data showed that the above medium effects are stronger for central collisions as compared to peripheral 
ones~\cite{Abelev:2014uua}. Here also we find that the description of the $p_T$ spectra improves as we go to more 
peripheral cases. In this regard it is to be noted that the lifetime of $\phi$ is an order of magnitude greater 
than that of $K^*$ and hence we manage to describe its yield and spectra well without incorporating the medium 
effects of rescattering and regeneration.

   \begin{figure}[htb]
  \begin{center}
    \scalebox{0.5}{\includegraphics{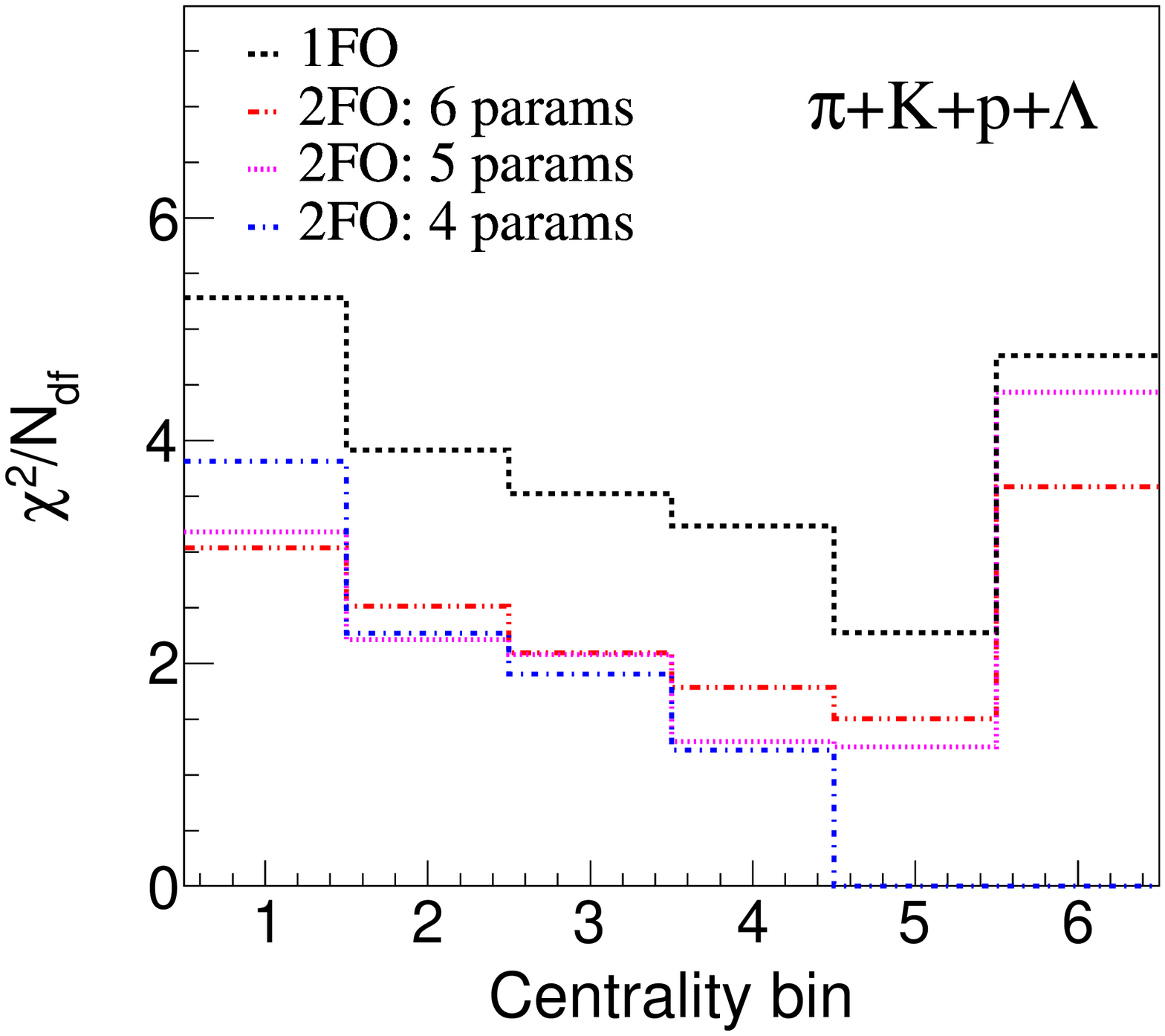}} 
  \caption{(Color online) Comparison of $\chi^2/N_{df}$ for different parameter sets in 2FO with 1FO. The last two peripheral 
  bins in case of the fit with 2FO and 4 parameters are not shown as $\frac{\tau_{ns}}{\tau_s}=1.3$ does not hold for these 
  centralities. For the most peripheral bin 6, $\l60-80\r\%$, the model fits are becoming worse indicating that the model should be
improved for peripheral collisions. 
}
  \label{fig.chi2lesspara}
  \end{center}
  \end{figure}

We close this section with a short discussion on the minimum number of free parameters required for a good 
description of the spectra data within the 2FO scheme. So far we have presented results for the most general 
case treating the thermal ($T$) as well as the geometric parameters ($\rho$ and $\tau$) for both the strange and 
non-strange freezeout surfaces as free parameters. Thus, we have used six free parameters. A closer look at the 
values of the extracted parameters as given in Table~\ref{tab.2FOparams} reveals interesting trends. We find 
the ratio of the geometric parameters of the fireball $\rho/\tau$ to be invariant under expansion from the 
strange to the non-strange FO surface
\beq
\l\frac{\rho_s}{\tau_s}\r=\l\frac{\rho_{ns}}{\tau_{ns}}\r
\label{eq.rhobytau}
\eeq
Thus very good description of the spectra data with similar $\chi^2/N_{df}$ is obtained with just five parameters 
using Eq.~\ref{eq.rhobytau} to fix $\rho_{ns}$. Further, we also note that except for the most peripheral centralities 
of $\l40-60\r\%$ and $\l60-80\r\%$, the ratio of $\tau_s$ and $\tau_{ns}$ is a constant across centralities. Thus the 
condition of $\frac{\tau_{ns}}{\tau_s}=1.3$ along with Eq.~\ref{eq.rhobytau} allows for a good 4 parameter (offcourse 
in order to determine the ratio $\tau_{ns}/\tau_s$, atleast one central bin needs to be fitted with five parameters) fit of the 
spectra data for the $\l0-40\r\%$ centralities data. In Fig.~\ref{fig.chi2lesspara}, we have compared the $\chi^2/N_{df}$ 
for different centralities in the 2FO framework with these different fitting schemes. The most general six parameter fit 
has been compared with the five parameter fit for all centralities and we find similar goodness of fit. Finally for the 
$\l0-40\r\%$ centralities, we have also shown the results from the four parameter fit which also yields a similar 
$\chi^2/N_{df}$. The corresponding 1FO values are also shown for comparison.

\section{Summary}\label{sec.summary}

   \begin{figure}[htb]
  \begin{center}
    \scalebox{0.6}{\includegraphics{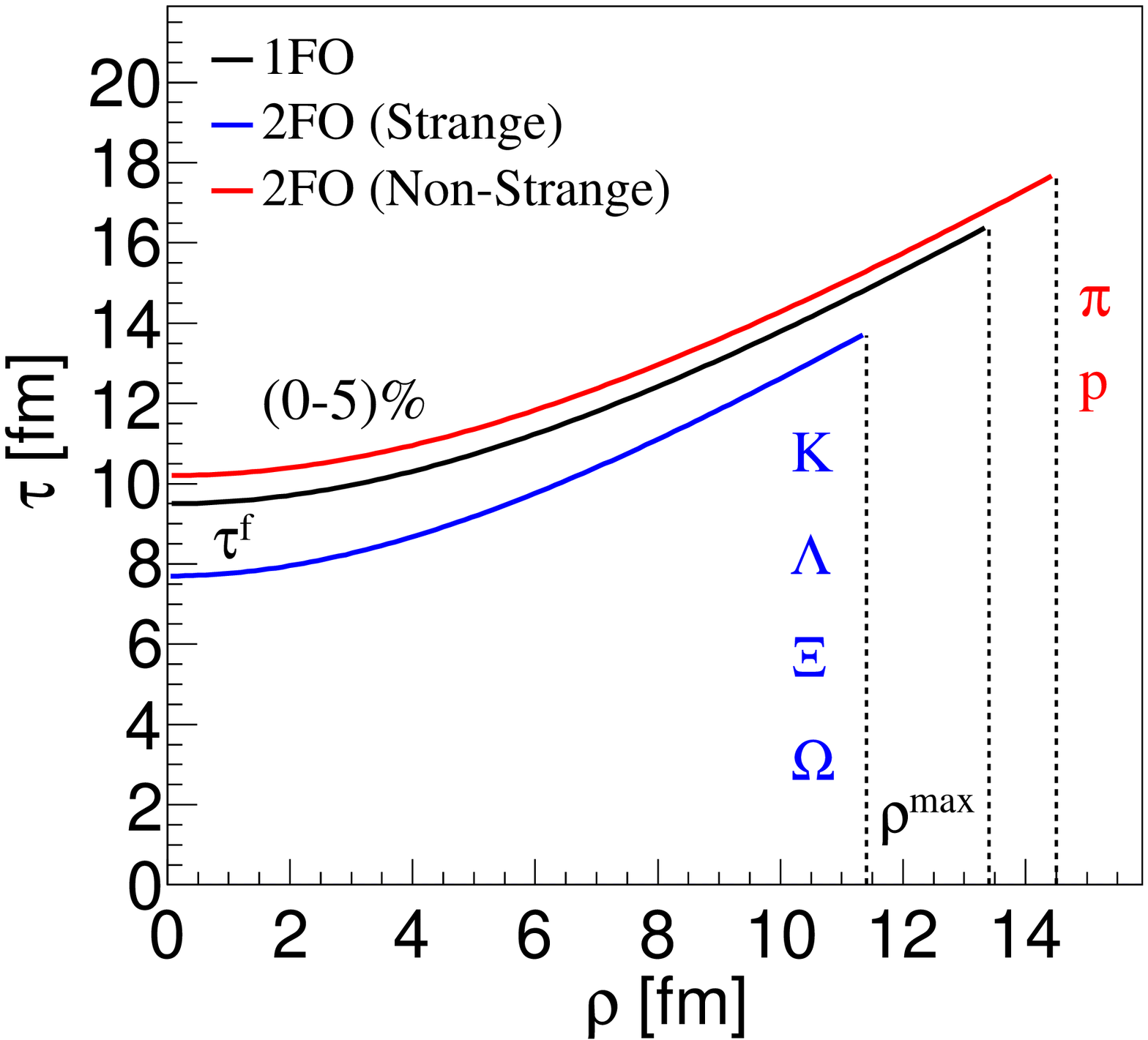}} 
  \caption{(Color online) The non-strange and strange freezeout hypersurfaces as obtained in the 1FO and 2FO 
  schemes from fits to the data on $p_T$ spectra for the $\l0-5\r\%$ centrality bin. The freezeout is assumed to 
  occur at a fixed invariant time $\tauf$ and $\rhom$ is the edge of the freezeout cylinder.
}
  \label{fig.freezeoutsurface}
  \end{center}
  \end{figure}

We have obtained the $p_T$ spectra of $\pi^+$, $K^+$, $p$, $\Lambda$, $\Xi^-$, $\Omega$, $K^*$ and $\phi$ within two 
schemes of freezeout: 1FO and 2FO. Overall, the description of the $p_T$ spectra improves considerably in 2FO as 
compared to 1FO with a reduction in $\chi^2/N_{df}$ by around $40\%$. The extracted values of the parameters for the 
strange and non-strange FHs show interesting correlations: $\frac{\rho_s}{\tau_{s}}=
\frac{\rho_{ns}}{\tau_{ns}}$ for all centralities and $\frac{\tau_{ns}}{\tau_s}=1.3$ for $\l0-40\r\%$ centralities. 
These conditions allow us to reduce the number of free parameters from six, in case of the most general 2FO scenario, 
to four with similar $\chi^2/N_{df}$. The FHs as obtained in the 1FO and 2FO schemes for the most 
central bin $\l0-5\r\%$ is plotted in Fig.~\ref{fig.freezeoutsurface}. A clear separation between 
the non-strange and strange FHs is visible with an earlier freezeout of the hadrons with non-zero 
strangeness content. As expected from the fitted values of $T$, $\rhom$ and $\tauf$, the 1FO hypersurface lies in between 
the non strange and strange FHs of 2FO. There are some indications of further substructures in 
the simple 2FO scheme that we have studied here. These are: (a) tension between the model spectra and data of $\pi$ and 
$p$ and (b) overprediction of the spectra in the case of multi-strange baryons. These could be hints of separate 
freezeout of $\pi$ and $p$ and early freezeout of the multi-strange baryons. However, before drawing any conclusion on 
the nature of freezeout based on these observations, it will be important to account for the contribution to the hadron 
spectra from the heavier resonances which are yet unobserved but predicted by quark models and hinted upon by LQCD results. 
Thus, while the current work based on 2FO with an early freezeout of the hadrons with non-zero strangeness content show considerable 
improvement over the conventional 1FO scheme in describing the data on transverse momentum spectra across all centralities, 
there are suggestions of possible further substructures in freezeout which are yet to be confirmed.

\section{Acknowledgement}
SC acknowledges discussions and collaborations on freezeout with Rohini Godbole and Sourendu Gupta and thanks ``Centre for 
Nuclear Theory" [PIC XII-R$\&$D-VEC-5.02.0500], Variable Energy Cyclotron Centre for support. BM and RS acknowledges 
financial support from DST SwarnaJayanti and  DAE-SRC projects of BM.

\bibliographystyle{apsrev4-1}
\bibliography{2CFO_spectra}

\end{document}